\documentclass[11pt,a4paper]{article}
\pdfoutput=1
\usepackage[utf8]{inputenc}
\usepackage{mystyle}
\usepackage{afterpage}
\usepackage{xcolor}
\usepackage{epsfig}
\usepackage{cancel}
\usepackage{framed}
\usepackage{multirow}

\usepackage{seqsplit}
\usepackage{enumitem} 

\usepackage{amsmath, amsthm, amsfonts, amssymb, latexsym}
\usepackage{upgreek}
\usepackage[caption=false]{subfig}

\usepackage{url}
\usepackage{ifpdf}

\def\beq{\begin{equation}}
\def\eeq{\end{equation}}

\def\logbar{\overline{\log\,}}

\def\bequ{\begin{equation}}
\def\eequ{\end{equation}}

\title{All one-loop scalar vertices in the effective potential approach}

\author[a]{Jos\'e~Eliel Camargo-Molina,}
\emailAdd{Eliel@thep.lu.se}
\author[a,b]{Ant\'onio P. Morais,} 
\emailAdd{aapmorais@ua.pt}
\author[a]{Roman Pasechnik,} 
\emailAdd{Roman.Pasechnik@thep.lu.se}
\author[b]{Marco~O.~P.~Sampaio,} 
\emailAdd{msampaio@ua.pt}
\author[a]{Jonas Wess\'en}
\emailAdd{Jonas.Wessen@thep.lu.se}

\affiliation[a]{Department of Astronomy and Theoretical Physics,
Lund University, \\ SE 223-62 Lund, Sweden} 
\affiliation[b]{Departamento de F\'\i sica da Universidade de Aveiro and CIDMA \\ 
Campus de Santiago, 3810-183 Aveiro, Portugal}


\abstract{Using the one-loop Coleman-Weinberg effective potential, we derive a general analytic expression for all the derivatives of the effective potential with respect to any number of classical scalar fields. The result is valid for a renormalisable theory in four dimensions with any number of scalars, fermions or gauge bosons. This result corresponds to the zero-external momentum contribution to a general one-loop diagram with $N$ scalar external legs. We illustrate the use of the general result in two simple scalar singlet extensions of the Standard Model, to obtain the dominant contributions to the triple couplings of light scalar particles under the zero external momentum approximation.   

}

\begin{document}

\maketitle

\section{Introduction}
\label{sec:Introduction}

The Large Hadron Collider (LHC) ATLAS~\cite{Aad:2012tfa} and CMS~\cite{Chatrchyan:2012ufa} experiments, have achieved in recent years a landmark in the history of particle physics: the direct confirmation of the first known fundamental scalar in Nature. Scalar fields are frequently used in the modelling of new physics beyond the Standard Model (SM) of particle physics and in cosmology. Interesting examples are models with scalar dark matter candidates~\cite{Silveira:1985rk, McDonald:1993ex, Burgess:2000yq, Bento:2000ah, Davoudiasl:2004be,Kusenko:2006rh, vanderBij:2006ne, He:2008qm, Gonderinger:2009jp, Mambrini:2011ik,He:2011gc, Gonderinger:2012rd, EliasMiro:2012ay, Cline:2013gha, Gabrielli:2013hma,Profumo:2014opa}, models of scalar field inflation~\cite{Guth:1980zm,Linde:1981mu,Guth:1982ec}, or even in attempts to explain the finer structure of the SM parameters through scalar flavour models\footnote{For a review in the context of neutrino physics see for example~\cite{King:2013eh}.}. Furthermore they are very often predicted as part of many Beyond the Standard Model (BSM) scenarios such as supersymmetric theories~\cite{Martin:1997ns}, models with extra dimensions~\cite{Antoniadis:1990ew,ArkaniHamed:1998rs}, little Higgs models~\cite{ArkaniHamed:2001nc} and grand unified theories~\cite{Dimopoulos:1981yj,Ibanez:1981yh,Ellis:1990wk}, among many others. Searches for scalar particle states are therefore of great interest and one of the focuses of the ongoing LHC searches. 

Whether the current 13~TeV run of the LHC provides us a new scalar state, such as the hypothetical state recently hinted for with a $750$~GeV mass~\cite{excessATLAS,CMS:2015dxe}, or whether it provides measurements of the SM Higgs self couplings, the low energy observables of these lighter scalars may receive contributions from new heavy states through radiative corrections. If the typical scale of the external momentum of the light particles involved in such observables is small compared to the masses of the hypothetical heavy particles inside the diagrams, then their contributions can be computed in the zero-external momentum approximation~\cite{Ellis:1990nz}. In that approximation, for the particular case of purely scalar operators with no derivative terms, such ``heavy'' contributions to the corresponding loop corrected vertices can be extracted solely from derivatives of the scalar effective potential. 

The computation of the scalar effective potential for a generic Quantum Field Theory (QFT), is often performed in dimensional regularisation and mass independent renormalisation schemes such as the $\overline{\mathrm{MS}}$ or $\overline{\mathrm{DR}}$ schemes for simplicity. This has been reviewed at two-loop order in such schemes and in the Landau gauge in~\cite{Martin:2001vx}. In this paper we start from the one-loop Coleman-Weinberg (CW) effective potential~\cite{Coleman:1973jx} for a generic QFT and extend the analysis to find a general analytic expression for all its derivatives (corresponding to an arbitrary number of external legs). This is done by simplifying the derivatives of the matrix-$\log$ of the field dependent mass squared matrix. It reduces to a combinatoric problem involving the tree level vertices of the theory in the tree level mass eigenbasis, and a set of totally symmetric tensors that are functions only of the physical masses of the eigenstates. Our central result consists of explicit expressions for the derivatives of the one-loop effective potential with an arbitrary number of derivatives. These expressions can be easily implemented and evaluated for any theory, either as analytic expressions or numerically (reducing to common linear algebra operations).  

Though our results focus on the derivatives of the effective potential, recently there have been advances in obtaining the functional derivatives of the one-loop effective action after integrating out heavy degrees of freedom using a covariant derivative expansion~\cite{Drozd:2015rsp,Henning:2014wua,delAguila:2016zcb,Henning:2016lyp}. In such approach all the operators can be obtained in addition to the translation invariant operators captured by the one-loop effective potential. Nevertheless, the results have to be computed at a fixed order in the derivative expansion while our general result for the effective potential contribution is valid for an arbitrary number of scalar field derivatives.

Finally, we also apply our results in two simple scalar singlet extensions of the SM that are still phenomenologically viable. We use our results to illustrate the importance of the Next to Leading Order (NLO) radiative corrections due to a new heavy scalar, to the triple vertices in the real (RxSM) and the complex (CxSM) singlet extensions of the SM recently analysed in~\cite{Costa:2015llh}.

The structure of the paper is the following. In Sect.~\ref{sec:conventions} we set conventions for a general QFT and define the various fields and respective couplings.~In Sect.~\ref{sec:OneLoopNpoint} we derive our main result by applying derivatives to the CW effective potential, where many of the technical steps are described in detail in appendix~\ref{app:Matrix_Log}. We also explicitly verify the symmetry properties of the result under the interchange of scalar indices in the remaining sub-sections for cases with up to four derivatives. In Sect.~\ref{sec:examples} we provide examples of applications and in Sect.~\ref{sec:conclusions} we summarise our conclusions.

\section{Notations and definitions}\label{sec:conventions} 
In this article we follow the general approach of~\cite{Martin:2001vx}, and write the most general Gauged QFT Lagrangian with fields of spin up to 1. Before choosing the vacuum of the theory, i.e. before symmetry breaking, we use the following notations for the various fields\footnote{The kinetic terms of the various fields are canonically normalised}:
\begin{itemize}
\item {\em Scalars}: All scalar multiplets are decomposed as $N_0$ real scalar fields, $\Phi_i$ with $i=1,\ldots,N_0$.
\item {\em Fermions}: All fermion multiplets are decomposed as $N_{1/2}$ Weyl 2-spinors, $\Psi_I$ with $I=1,\ldots,N_{1/2}$.
\item {\em Gauge bosons} are represented by a 4-vector with a gauge group index running over $N_1$ bosons in the adjoint representation of the gauge group, i.e. $\mathcal{A}_a^{\mu}$. 
\end{itemize}
The most general renormalisable interaction Lagrangian involving the scalar sector\footnote{We suppress the interaction terms without scalar fields because we will focus on the one-loop effective potential.} is then written as:
\begin{eqnarray}
-\mathcal{L}_{S} & = & L^{i}\Phi_{i}+\dfrac{1}{2!}L^{ij}\Phi_{i}\Phi_{j}+\dfrac{1}{3!}L^{ijk}\Phi_{i}\Phi_{j}\Phi_{k}+\dfrac{1}{4!}L^{ijkl}\Phi_{i}\Phi_{j}\Phi_{k}\Phi_{l}\; ,\nonumber\\
-\mathcal{L}_{F} & = & \frac{1}{2}Y^{IJ}\Psi_{I}\Psi_{J}+\frac{1}{2}Y^{IJk}\Psi_{I}\Psi_{J}\Phi_{k}+ {\rm c.c.} \; ,\label{Eq:L-basis}\\
-\mathcal{L}_{SG} & = & \dfrac{1}{4}G^{abij}\mathcal{A}_{a\mu}\mathcal{A}_{b}^{\mu}\Phi_{i}\Phi_{j}+G^{aij}\mathcal{A}_{a\mu}\Phi_{i}\partial^{\mu}\Phi_{j}\; .\nonumber
\end{eqnarray}
We adopt the Einstein convention where repeated indices that are one up (superscript) and one down (subscript) are summed over. We also adopt the convention that repeated indices that are all down or all up are \underline{not} summed over. This will become useful to define tensor components in the mass eigenstate basis. The scalar couplings in the gauge eigen-basis are denoted by $\left\{L^i,L^{ij},L^{ijk},L^{ijkl}\right\}$ respectively for the linear, quadratic, cubic and quartic couplings and they are totally symmetric in the scalar indices $i,j,k,l$. In this decomposition, they are all real.\footnote{We follow the notation in~\cite{Costa:2014qga} where basis invariant expressions for the two-loop beta functions of the scalar couplings were derived using the two-loop effective potential.} The fermion quadratic and Yukawa terms are denoted by the complex numbers $Y^{IJ}$ and $Y^{IJk}$, which are symmetric under interchange of fermionic indices. The gauge-scalar couplings are denoted by $G^{abij}$ and $G^{aij}$ (see also~\cite{Martin:2001vx} for details). 

After symmetry breaking, the scalar fields are shifted around a classical field configuration. We denote a generic classical field configuration, around which the perturbative calculations are set, by $v_i$ and shift the fields as follows: $\Phi_{i}(x)=v_{i}+\phi_{i}(x)$ (now $\phi_i$ are the quantum scalar field fluctuations around the classical configuration $v_i$). Then we obtain the following Lagrangian in a basis that we define to be the $\Lambda$-basis\footnote{Note that the Landau gauge conditions can be used to eliminate a term $G^{aij}v_{i}\mathcal{A}_{a\mu}\partial^{\mu}\Phi_{j}$, up to a surface term, through an integration by parts.} 
\begin{eqnarray}
-\mathcal{L}_{S} & = & \Lambda+\Lambda^{i}_{(S)}\phi_{i}+\dfrac{1}{2}\Lambda^{ij}_{(S)}\phi_{i}\phi_{j}+\dfrac{1}{3!}\Lambda^{ijk}_{(S)}\phi_{i}\phi_{j}\phi_{k}+\dfrac{1}{4!}\Lambda^{ijkl}_{(S)}\phi_{i}\phi_{j}\phi_{k}\phi_{l}\; ,\nonumber\\
-\mathcal{L}_{F} & = & \frac{1}{2}M^{IJ}\Psi_{I}\Psi_{J}+\frac{1}{2}Y^{IJk}\Psi_{I}\Psi_{J}\phi_{k}+{\rm c.c.}\;, \label{Eq.Lambda-basis}\\
-\mathcal{L}_{SG} & = & \dfrac{1}{2}\Lambda_{(G)}^{ab}\mathcal{A}_{a\mu}\mathcal{A}^{\mu}_b+\dfrac{1}{2}\Lambda^{abi}_{(G)}\mathcal{A}_{a\mu}\mathcal{A}_{b}^{\mu}\phi_{i}+\dfrac{1}{4}\Lambda^{abij}_{(G)}\mathcal{A}_{a\mu}\mathcal{A}_{b}^{\mu}\phi_{i}\phi_{j}+G^{aij}\mathcal{A}_{a\mu}\phi_{i}\partial^{\mu}\phi_{j}\;,\nonumber
\end{eqnarray}
where 
\begin{eqnarray}
\Lambda & \equiv & L^{i}v_{i}+\dfrac{1}{2!}L^{ij}v_{i}v_{j}+\dfrac{1}{3!}L^{ijk}v_{i}v_{j}v_{k}+\dfrac{1}{4!}L^{ijkl}v_{i}v_{j}v_{k}v_{l}=V^{(0)}(v_i)\;,\nonumber\\
\Lambda^{i}_{(S)} & \equiv & L^{i}+L^{ij}v_{j}+\dfrac{1}{2}L^{ijk}v_{j}v_{k}+\dfrac{1}{6}L^{ijkl}v_{j}v_{k}v_{l}\;,\nonumber\\
\Lambda_{(S)}^{ij} & \equiv & L^{ij}+L^{ijk}v_{k}+\dfrac{1}{2}L^{ijkl}v_{k}v_{l}\;, \label{Eq:lambda_defs1} \\
\Lambda^{ijk}_{(S)} & \equiv & L^{ijk}+L^{ijkl}v_{l}\;,\nonumber\\
\Lambda^{ijkl}_{(S)} & \equiv & L^{ijkl} \nonumber\;,
\end{eqnarray}
 and 
\begin{eqnarray}
\Lambda_{(G)}^{ab} & \equiv & \dfrac{1}{2}G^{abij}v_{i}v_{j}\;,\nonumber\\
\Lambda^{abi}_{(G)} & \equiv & G^{abij}v_{j}\;,\label{Eq:lambda_defs2}\\
\Lambda^{abij}_{(G)} & \equiv & G^{abij} \nonumber \; .
\end{eqnarray}
 We have defined in Eqs.~\eqref{Eq:lambda_defs1} and~\eqref{Eq:lambda_defs2} the scalar mass-squared matrix, $\Lambda_{(S)}^{ij}$, the gauge boson mass-squared matrix $\Lambda_{(G)}^{ab}$ and the tree level effective potential $V^{(0)}(v_i)$. The fermion mass-squared matrix is obtained from 
\begin{equation}\label{eq:F_MassSquared}
\Lambda_{(F)}^{IJ} \equiv  M^{*IL}M_{L}^{\phantom{L}J} =  Y^{*IL}Y_{L}^{\phantom{L}J}+\left(Y^{JL}Y_{\phantom{*}L}^{*\phantom{L}Ik}+Y^{*IL}Y_{L}^{\phantom{L}Jk}\right)v_{k}+Y^{*ILk}Y_{L}^{\phantom{L}Jm}v_{k}v_{m}\;,
\end{equation}
where we use the fermion mass matrix defined through the shifted Lagrangian, Eq.~\eqref{Eq.Lambda-basis}:
\begin{equation}
M^{IJ}=Y^{IJ}+Y^{IJk}v_{k} \; .
\end{equation}
 Note that all (non-spacetime) Latin indices are assumed to be in Euclidean
space (they are lowered and raised with the identity matrix). Another important point is that, though the fermionic tensors that appear directly in the Lagrangian are symmetric under interchange of fermionic indices, the fermion mass-squared matrix defined in Eq.~\eqref{eq:F_MassSquared} is not necessarily symmetric. In general it is, however, hermitian. Below we will define fermionic cubic and quartic effective vertices that are also hermitian with respect to the two fermionic indices.

 This basis will be particularly useful to obtain expressions for the derivatives of the effective potential analytically, which are directly related to the scalar $N$-point functions of the theory at zero external momenta. The final results, however, will adopt a much simpler form in a third basis, which we name as the $\lambda$-basis. This is defined as the basis that diagonalises all the tree level mass-squared matrices. In the $\lambda$-basis the Lagrangian takes the form
\begin{eqnarray}
-\mathcal{L}_{S}	&=&	\Lambda+\lambda^{i}_{(S)}R_{i}+\dfrac{1}{2}m^{2\;\;i}_{(S)}R_{i}^{2}+\dfrac{1}{3!}\lambda^{ijk}_{(S)}R_{i}R_{j}R_{k}+\dfrac{1}{4!}\lambda^{ijkl}_{(S)}R_{i}R_{j}R_{k}R_{l} \;,\nonumber\\
-\mathcal{L}_{F}	&=&	\frac{1}{2}m^{IJ}\psi_{I}\psi_{J}+\frac{1}{2}y^{IJk}\psi_{I}\psi_{J}R_{k}+{\rm c.c.} \;,\label{Eq:lambda_basis} \\
-\mathcal{L}_{SG}	&=&	\dfrac{1}{2}m^{2\;\;a}_{(G)}A_{a\mu}A_{a}^{\mu}+\dfrac{1}{2}\lambda^{abi}_{(G)}A_{a\mu}A_{b}^{\mu}R_{i}+\dfrac{1}{4}\lambda^{abij}_{(G)}A_{a\mu}A_{b}^{\mu}R_{i}R_{j}+\lambda^{aij}_{(G)}A_{a\mu}R_{i}\partial^{\mu}R_{j} \;,\nonumber
\end{eqnarray}  
where the rotated fields are 
\begin{eqnarray}
R_{i}&=&\left[O_{(S)}\right]_{\phantom{j}i}^{j}\phi_{j}\nonumber \\
A_{a\mu}&=&\left[O_{(G)}\right]_{\phantom{b}a}^{b}\mathcal{A}_{b\mu}\label{Eq:Rot-fields}\\
\psi_{I}&=&\left[U_{(F)}^{*}\right]_{\phantom{J}I}^{J}\Psi_{J} \; . \nonumber
\end{eqnarray}
Here we must use orthogonal matrices for the bosonic rotations and unitary matrices for the fermions. If we denote, generically, such a transformation matrix for the field of type $T=\{S,G,F\}$ by $\mathcal{U}_{(T)}$ (unitary or orthogonal) then all mass-squared matrices are diagonal in this basis so the $\Lambda$-basis mass-squared matrices defined above obey 
\begin{equation}
\mathcal{U}_{(T)}\Lambda_{(T)}\mathcal{U}_{(T)}^\dagger={\rm diag}\{m_{(T)a}^2\} \; .
\end{equation}
Note that on the right hand side we have now used Latin indices from the beginning of the alphabet to denote the component of the diagonal. Whenever the type $T$ is not specified, we follow the convention of using lower case indices from the beginning of the Latin alphabet ($a,b,c,\ldots$) and reserve  indices from the middle of the alphabet ($i,j,k,\ldots$) for scalar field indices. Finally, note that all couplings in the $\lambda$-basis, Eq.~\eqref{Eq:lambda_basis}, are now in lower case. The transformation relating them to the corresponding upper case couplings defined in Eq.~\eqref{Eq.Lambda-basis} ($\Lambda$-basis) is obtained by rotating each index using the $\mathcal{U}_{(T)}$ matrix corresponding to the index type ($S,F$ or $G$) as induced by the transformations in Eq.~\eqref{Eq:Rot-fields}.

In the following sections it will also be useful to note that the derivatives of the various mass-squared matrices with respect to the $v_i$ are related to the cubic and quartic couplings as follows: 
\begin{eqnarray}
 \partial^{k}\Lambda_{(S)}^{ij}	&=&	L^{ijk}+L^{ijkl}v_{l}=\Lambda^{ijk}_{(S)}\;,\nonumber \\
\partial^{kl}\Lambda_{(S)}^{ij}	&=&	L^{ijkl}=\Lambda^{ijkl}_{(S)} \;,\nonumber\\
 \partial^{k}\Lambda_{(F)}^{IJ}	&=&	\partial^{k}M^{*IL}M_{L}^{\phantom{L}J}+M^{*IL}\partial^{k}M_{L}^{\phantom{L}J} \;,
	=	Y^{*ILk}M_{L}^{\phantom{L}J}+M^{*IL}Y_{L}^{\phantom{L}Jk}\equiv\Lambda^{IJk}_{(F)} \;,\nonumber\\
\partial^{km}\Lambda_{(F)}^{IJ}	&=&	Y^{*ILk}\partial^{m}M_{L}^{\phantom{L}J}+\partial^{m}M^{*IL}Y_{L}^{\phantom{L}Jk}\;,
	=	Y^{*ILk}Y_{L}^{\phantom{L}Jm}+Y^{*ILm}Y_{L}^{\phantom{L}Jk}\equiv\Lambda^{IJkm}_{(F)} \;,\nonumber\\
 \partial^{i}\Lambda_{(G)}^{ab}	&=&	G^{abij}v_{j}=\Lambda^{abi}_{(G)}\;,\nonumber\\
\partial^{ij}\Lambda_{(G)}^{ab}	&=&	G^{abij}=\Lambda^{abij}_{(G)} \; . \label{Eq:Lambda-couplings-derivatives}
 \end{eqnarray}
We use the notation $\partial_i=\frac{\partial}{\partial v_i}$, $\partial_{ij}=\frac{\partial^2}{\partial v_i\partial v_j}$ etc... for derivatives with respect to the arbitrary classical scalar field configuration $v_i$.
If we denote generically the cubic and quartic couplings for type $T$ (on the right hand side of each of the equations in~\eqref{Eq:Lambda-couplings-derivatives})  by $\Lambda_{(T)abi}$ and $\Lambda_{(T)abij}$ respectively, then these relations can be written collectively as (after lowering the indices with the Euclidean metric)
\begin{eqnarray}
\partial_{i}\Lambda_{(T)ab}&=&\Lambda_{(T)abi}\;, \nonumber\\
\partial_{ij}\Lambda_{(T)ab}&=&\Lambda_{(T)abij} \; .
\end{eqnarray}

\section{One-loop $N$-point vertices at zero external momenta}\label{sec:OneLoopNpoint}
In this section we present the general analytic expressions for the $N$th order derivatives of the effective potential. We start from the general one-loop contribution to the effective potential in the Landau gauge that is given by~\cite{Martin:2001vx} 
\begin{equation}
V^{(1)}=\dfrac{1}{4}\sum_{T}(-1)^{2s_{T}}(1+2s_{T}){\rm Tr}\left[\Lambda_{(T)}^{2}\left(\logbar\Lambda_{(T)}-k_{T}\right)\right] \; .
\end{equation}
Here the general loop expansion of the effective potential is defined as 
\begin{equation}
V_{\rm eff}\equiv \sum_n\varepsilon^n V^{(n)}\; ,
\end{equation} with $\varepsilon=\hbar/(4\pi)^2$ so that $V^{(n)}$ is the $n$-loop effective potential.
The spin of the field is denoted by $s_{T}$ and we have defined $\logbar(m^2)\equiv \log(m^2/\mu^2)$ with $\mu$ the renormalisation scale and 
$k_{T}$ depends on the renormalisation scheme ($\overline{\rm MS}$ or $\overline{\rm DR}$) and can be specified later. The $\log$ function is defined over matrices. 

We now state the general result. Further details of the proof are provided in appendix~\ref{app:Matrix_Log}. Applying the $N$-th order derivative operator with respect to the fields $v_{i_1},\ldots,v_{i_N}$, in the $\Lambda$-basis, to the one-loop effective potential (with $N\geq 1$) we obtain 
\begin{multline}
\partial_{i_{1},\ldots,i_{N}}V^{(1)}=\partial_{i_{1},\ldots,i_{N-1}}\dfrac{1}{4}\sum_{T}(-1)^{2s_{T}}(1+2s_{T})\times \\
\times {\rm Tr} \left[\partial_{i_N}\left(\Lambda_{(T)}^{2}\right)\left(\logbar\Lambda_{(T)}-k_{T}\right)
+\Lambda_{(T)}^{2}\partial_{i_N}\left(\logbar\Lambda_{(T)}\right)\right]\;.
 \end{multline}
In appendix~\ref{app:Matrix_Log} we prove the result that ${\rm Tr}\left[\Lambda_{(T)}^{2}\partial_{i}\left(\logbar\Lambda_{(T)}\right)\right]=\frac{1}{2}{\rm Tr}\left[\partial_{i}\left(\Lambda_{(T)}^2\right)\right]$. Using this result, we can effectively lower by one the order of the derivatives acting on the $\logbar$. Using this identity, the cyclic property of the trace and the fact that a function of a matrix commutes with the matrix itself we obtain
\begin{multline}
\partial_{i_{1},\ldots,i_{N}}V^{(1)}=\partial_{i_{1},\ldots,i_{N-1}}\sum_{T}\tfrac{(-1)^{2s_{T}}(1+2s_{T})}{4} {\rm Tr} \left[\partial_{i_N}\left(\Lambda_{(T)}^{2}\right)\left(\logbar\Lambda_{(T)}-k_{T}+\frac{1}{2}\right)\right]\; .
 \end{multline}
So we have reduced by one the order of the derivative that will act on the matrix-$\logbar$\footnote{Note however that the matrix multiplying the $\logbar$ now contains derivatives, so it no longer commutes with $\Lambda_{(T)}$ and this process cannot be continued.}. Acting with the remaining $N-1$ derivatives we obtain an expression in the form
\begin{multline}\label{Eq:n-point}
\partial_{i_{1},\ldots,i_{N}}V^{(1)}=\dfrac{1}{4}\sum_{T}(-1)^{2s_{T}}(1+2s_{T}) \times \\
{\rm S}_{\left\{ i_{1}\ldots i_{N-1}\right\} }\left[{\rm Tr}\sum_{p=0}^{\min\{N-1,3\}}\binom{N-1}{p}\partial_{i_{1},\ldots,i_{p},i_{N}}^{(p+1)}\left(\Lambda_{(T)}^{2}\right)\partial_{i_{p+1},\ldots,i_{N-1}}\left(\logbar\Lambda_{(T)}-k_{T}+\tfrac{1}{2}\right)\right]\;,
 \end{multline}
where the innermost sum is over all possible partitions of the $\{i_1,\ldots,i_{N-1}\}$ indices in two lists and we define here the operator ${\rm S}_{\left\{ i_1\ldots i_{N-1}\right\} }$, which denotes symmetrisation with respect to the $(N-1)$-indices in the list. Observe that in the range of the sum over $p$ we have used the fact that the mass-squared matrix only has non-zero derivatives up to order two. The term in the product that contains derivatives acting on the square of the mass-squared matrix is simple to obtain (since it is polynomial). For notational simplicity we present the result making use of the following set of tensors
\begin{equation}\label{eq:sigma_def}
\sigma^{(p)}_{(T)a b i_1\ldots i_p}\equiv \mathcal{U}_{(T)a}^{\phantom{(T)a}c}\,\mathcal{U}_{(T)\phantom{.}b}^{\dagger \phantom{...}d}O_{(S)\phantom{j_1}i_1}^{\phantom{(S)}j_1}\ldots O_{(S)\phantom{j_N}i_p}^{\phantom{(S)} j_p}\left[\partial_{j_{1}\ldots j_{p}}\left(\Lambda_{(T)}^2\right)_{cd}\right]\; .
\end{equation}
The components for the non-zero cases are
\begin{equation} \label{eq:PartialPoly}
\begin{cases}
\sigma^{(1)}_{(T)a b i}=m_{(T)a}^{2}\lambda_{(T)abi}+\lambda_{(T)abi}m_{(T)b}^{2}\;,\vspace{2mm}\\
\sigma^{(2)}_{(T)a b i j}=m_{(T)a}^{2}\lambda_{(T)abij}+\lambda_{(T)abij}m_{(T)b}^{2}+\lambda_{(T)aci}\lambda_{(T)\phantom{c}bj}^{\phantom{(T)}c}+\lambda_{(T)acj}\lambda_{(T)\phantom{c}bi}^{\phantom{(T)}c}\;,\vspace{2mm}\\
\sigma^{(3)}_{(T)a b i j k}=\lambda_{(T)acij}\lambda_{(T)bk}^{c}+\lambda_{(T)bcij}\lambda_{(T)ak}^{c}+(j\leftrightarrow k)+(i\leftrightarrow k)\;,\vspace{2mm}\\
\sigma^{(4)}_{(T)a b i j k l}=\lambda_{(T)acij}\lambda_{(T)bkl}^{c}+\lambda_{(T)bcij}\lambda_{(T)akl}^{c}+(j\leftrightarrow k)+(i\leftrightarrow k) \;.&
\end{cases}
\end{equation}
Here the cubic and quartic couplings for field of type $T$ in the $\lambda$-basis are generically denoted by $\lambda_{(T)abi}$ and $\lambda_{(T)abij}$. They are obtained from the corresponding $\Lambda_{(T)abi}$ and $\Lambda_{(T)abij}$ couplings using the $\mathcal{U}_{(T)}$ transformation matrices. For the second term in the product, Eq.~\eqref{Eq:n-point}, we define the following tensor: 
\begin{equation}\label{eq:delta_def}
\delta^{(N)}_{(T)a b i_1\ldots i_N}\equiv \mathcal{U}_{(T)a}^{\phantom{(T)a}c}\,\mathcal{U}_{(T)\phantom{.}b}^{\dagger \phantom{...}d}O_{(S)\phantom{j_1}i_1}^{\phantom{(S)}j_1}\ldots O_{(S)\phantom{j_N}i_N}^{\phantom{(S)} j_N}\left[\partial_{j_{1}\ldots j_{N}}\left(\logbar\Lambda_{(T)}-k_{T}+\frac{1}{2}\right)_{cd}\right]\;,
\end{equation}
which, for $N>0$, is a rotated version of the derivatives of the matrix-$\logbar$. In appendix~\ref{app:Matrix_Log} we derive a general algebraic analytic expressions for the $\delta^{(N)}_{a b i_1\ldots i_N}$ tensors. The general expression, Eq.~\eqref{eq:Gen_result}, contains another set of useful tensors whose components are
\begin{equation}\label{eq:fT-tensors}
f^{(k)}_{(T)a_{1}\ldots a_{N}}\equiv \sum_{x=1}^{N}\dfrac{{m^{2k}_{(T)}}_{a_x} \logbar {m^2_{(T)}}_{a_x}}{\prod_{y\neq x}\left(m^2_{(T)a_x}-m^2_{(T)a_y}\right)}\; .
\end{equation}
More explicitly, the first few cases that contribute to the 2, 3 and 4-point functions are (using Eqs.~\eqref{eq:N1case},~\eqref{eq:N2case} and~\eqref{eq:Gen_result}),
\begin{eqnarray}
 \delta^{(0)}_{(T)a b}&=&\delta_{ab}\left(\logbar m_{(T)a}^2-k_{T}+\tfrac{1}{2}\right), \nonumber\\
 \delta^{(1)}_{(T)a b i}&=&f^{(0)}_{(T)ab}\lambda_{(T)a b i}, \nonumber\\
 \delta^{(2)}_{(T)a b i j}&=&f^{(0)}_{(T)abc}\left(\lambda_{(T)a\phantom{c}i}^{\phantom{(T)a}c}\lambda_{(T)\phantom{j}b j}^{\phantom{(T)}c}+\lambda_{(T)a\phantom{c} j}^{\phantom{(T)a}c}\lambda_{(T)\phantom{j}b i}^{\phantom{(T)}c}\right)+f^{(0)}_{(T)ab}\lambda_{(T)ab i j}^{\phantom{(T)ci_{1i_{1}}}}\;, \label{eq:N1N2N3N4} \\
 \delta^{(3)}_{(T)a b i j k}&=& 3{\rm S}_{\left\{ i j k\right\} }\left[ 2\, f^{(0)}_{(T)acdb}\lambda_{(T)a\phantom{c}i}^{\phantom{(T)}\phantom{a}c}\lambda_{(T)\phantom{cd}j}^{\phantom{(T)}cd}\lambda_{(T)\phantom{d}bk}^{\phantom{(T)}d}+\right. \nonumber\\
&&\left.\phantom{2\, f^{(0)}_{(T)acdb}\lambda_{(T)a\phantom{c}i}^{\phantom{(T)}\phantom{a}c}\lambda_{(T)\phantom{cd}j}^{\phantom{(T)}cd}\lambda_{(T)\phantom{d}bk}^{\phantom{(T)}d}}+f^{(0)}_{(T)acb}\left(\lambda_{(T)a\phantom{c}ij}^{\phantom{(T)}\phantom{a}c}\lambda_{(T)\phantom{c}bk}^{\phantom{(T)}c}+\lambda_{(T)a\phantom{c}i}^{\phantom{(T)a}c}\lambda_{(T)\phantom{c}bjk}^{\phantom{(T)}c}\right)\right] \;. \nonumber
\end{eqnarray}
 We emphasise here that the repeated indices are summed over only once and that all repeated indices are all inside the same (suppressed) sum symbol according to the Einstein convention -- the fact that several repeated indices appear simultaneously up and down is a peculiarity of the mass eigenbasis (or diagonal basis). 

 We finally present the general result for the derivatives of the one-loop effective potential. This is fully determined using the $\delta^{(N)}_{(T)ab,i_1\ldots i_N}$ and $\sigma^{(N)}_{(T)ab,i_1\ldots i_N}$ tensors and we obtain
\begin{multline}\label{Eq.final-npoint}
\partial_{i_{1},\ldots,i_{N}}V_{\rm eff}=\Lambda_{i_{1},\ldots,i_{N}}^{(N\leq 4)}+\varepsilon O_{(S)\phantom{jj}i_{1}}^{\phantom{(S)}j_{1}}\ldots O_{(S)\phantom{jj}i_{N-1}}^{\phantom{(S)}j_{N-1}}O_{(S)\phantom{jj}i_{N}}^{\phantom{(S)}j_{N}}\sum_{T}(-1)^{2s_{T}}\frac{(1+2s_{T})}{4}\times \\
\times{\rm S}_{\left\{ j_{1}\ldots j_{N-1}\right\} }\left[\sum_{p=0}^{\min\{N-1,3\}}\binom{N-1}{p}\sigma_{(T)\phantom{p1ab}j_{1},\ldots,j_{p},j_{N}}^{(p+1)ab}\delta_{(T)baj_{p+1},\ldots,j_{N-1}}^{(N-1-p)}\right]+\mathcal{O}(\varepsilon^{2})\;.
 \end{multline} In the first term of Eq.~\eqref{Eq.final-npoint} we note that for $N>4$ the tree level term does not exist since we are working with a renormalisable theory. 

The final result, Eq.~\eqref{Eq.final-npoint}, may in principle contain infrared (IR) divergences. This is a well known issue of the effective potential~\cite{Elias-Miro:2014pca,Martin:2014bca} but it is also well known that the $p^2$ dependent contributions to the vertices must cancel out such IR divergences. In practice one may introduce an IR regulator whenever a massless particle is present in Eq.~\eqref{Eq.final-npoint} to safely identify and discard the IR divergences. Furthermore, typically the situation is even more favourable because: 
 i) for external lines with massless scalar states the IR divergences must cancel because the approximation $p^2\rightarrow 0$ for the external momenta becomes exact so the effective potential description is complete, and ii) in the examples that we will consider the only massless states are the SM-like Goldstone bosons and their couplings are such that, at one-loop, no IR divergences appear for triple vertices with the other physical scalars appearing in the external lines.

A final concern may be how to deal with particular limits with degenerate masses in Eq.~\eqref{eq:fT-tensors}. This is only an apparent issue. For all derivatives up to order four we will show, in the next subsections, that the result is always expressed in terms of $f^{(1)}$-tensors which are more regular than $f^{(0)}$ tensors. In addition expanding the $f^{(1)}$-tensors around the degenerate limit we find that, in fact, no extra divergences occur besides the IR ones. 

\subsection{First derivatives}

The results that we have found can be directly applied to obtain the first derivatives, which are usually necessary for the tadpole conditions that define the vacuum of the theory. Then, at one-loop order they are given by
\begin{eqnarray}
\partial_iV_{\rm eff}&=&\partial_iV^{(0)}+\varepsilon \partial_iV^{(1)}+\mathcal{O}(\varepsilon^{2}) \nonumber \\
&=&\Lambda_i+\varepsilon \partial_iV^{(1)}+\mathcal{O}(\varepsilon^{2})\;.
\end{eqnarray}
Using Eqs.~\eqref{eq:PartialPoly}, \eqref{eq:delta_def} and~\eqref{eq:N1N2N3N4}  in Eq.~\eqref{Eq:n-point} specialised to $N=1$, we obtain that the tadpole truncated at one-loop is
\begin{equation}
\partial_iV_{\rm eff}=\Lambda_i+\varepsilon O_{(S)\phantom{j}i}^{\phantom{(S)}j}\sum_{T}\tfrac{(-1)^{2s_{T}}(1+2s_{T})}{2}m_{(T)a}^{2}\lambda_{(T)a{\phantom a}j}^{\phantom{(T)a}a}\left(\logbar m_{(T)a}^{2}-k_{T}+\frac{1}{2}\right)+\mathcal{O}(\varepsilon^{2})\;.
\end{equation}

\subsection{Second derivatives}
The second derivatives are also straightforward to obtain in an explicit form. They are important to determine the one-loop correction to the masses of the particles. The zero external momentum contribution to the one-loop scalar two point function is obtained from 
\begin{eqnarray}
\partial_{kl}V_{\rm eff}&=&\Lambda_{kl}+\varepsilon O_{(S)\phantom{k}k}^{\phantom{(S)}i}O_{(S)\phantom{l}l}^{\phantom{(S)}j}\sum_{T}\tfrac{(-1)^{2s_{T}}(1+2s_{T})}{2}{\rm S}_{\{ij\}}\left[\lambda_{(T)i}^{ab}\lambda_{(T)baj}\left(f_{(T)ab}^{(1)}-k_{T}+\frac{1}{2}\right)+\right. \nonumber\\
&&\left.+\lambda_{(T)aij}^{a}m_{(T)a}^{2}\left(\logbar m_{(T)a}^{2}-k_{T}+\frac{1}{2}\right)\right]+\mathcal{O}(\varepsilon^2)
 \end{eqnarray}
where now we have used the identity
\begin{equation}
m_{a}^{2}f^{(0)}_{(T)ab_{1}\ldots b_{n}}=f_{(T)ab_{1}\ldots b_{n}}^{(1)}-f^{(0)}_{(T)b_{1}\ldots b_{n}}\delta_{aa} \; .
\end{equation}
This result can be used to evaluate the one-loop pole masses associated with the propagators of the scalar states in the theory, in the zero external momentum approximation. For the special case of massless states, such as for Goldstones, where the pole conditions are evaluated at zero external momentum, the second derivatives of the effective potential provide the exact pole conditions.
\subsection{Third and fourth derivatives}
Applying the same procedure one obtains higher derivatives. The third derivative is
\begin{eqnarray}
\partial_{lmn}V_{\rm eff}&=&\Lambda_{lmn}+\varepsilon O_{(S)\phantom{k}l}^{\phantom{(S)}i}O_{(S)\phantom{l}m}^{\phantom{S}j}O_{(S)\phantom{k}n}^{\phantom{(S)}k}\sum_{T}\tfrac{(-1)^{2s_{T}}(1+2s_{T})}{2}{\rm S}_{\{ijk\}}\left[2f_{(T)abc}^{(1)}\lambda_{(T)i}^{ab}\lambda_{(T)j}^{bc}\lambda_{(T)k}^{ca}+\right.\nonumber\\
&&\left.+3\lambda_{(T)ij}^{ab}\lambda_{(T)bak}\left(f_{(T)ab}^{(1)}-k_{T}+\tfrac{1}{2}\right)\right]+\mathcal{O}(\varepsilon^2)\;.  \label{Eq:3pointSimple}
 \end{eqnarray}
Note that, despite the apparent break down of the total symmetry under exchange of the scalar indices in the reduction leading to Eq.~\eqref{Eq:n-point}, the final result is manifestly symmetric as it should -- see  Eq.~\eqref{Eq:3pointSimple}. The same procedure can be applied for the fourth derivative to obtain
\begin{eqnarray}
\partial_{j_1j_2j_3j_4}V_{\rm eff}^{(1)}&=&\Lambda_{j_1j_2j_3j_4}+\varepsilon O_{(S)\phantom{k}j_1}^{\phantom{(S)}i_1}O_{(S)\phantom{k}j_2}^{\phantom{(S)}i_2}O_{(S)\phantom{k}j_3}^{\phantom{(S)}i_3}O_{(S)\phantom{k}j_4}^{\phantom{(S)}i_4}\times \nonumber \\
&&\times\sum_{T}\tfrac{(-1)^{2s_{T}}(1+2s_{T})}{2}3{\rm S}_{\{i_{1}i_{2}i_{3}i_{4}\}}\left[\lambda_{(T)i_{1}i_{4}}^{ab}\lambda_{(T)bai_{2}i_{3}}\left(f_{(T)ab}^{(1)}-k_{T}+\tfrac{1}{2}\right)+\right.\nonumber\\
&&\left.+2\left\{ f_{(T)abcd}^{(1)}\lambda_{(T)i_{4}}^{ab}\lambda_{(T)\phantom{c}i_{1}}^{\phantom{(T)}bc}\lambda_{(T)\phantom{cd}i_{2}}^{\phantom{(T)}cd}\lambda_{(T)\phantom{d}ai_{3}}^{\phantom{(T)}d}+2f_{(T)abc}^{(1)}\lambda_{(T)i_{4}}^{ab}\lambda_{(T)\phantom{c}i_{1}i_{2}}^{\phantom{(T)}bc}\lambda_{(T)\phantom{c}i_{3}}^{\phantom{(T)}ca}\right\} \right]\nonumber\\
&&+\mathcal{O}(\varepsilon^2)\;.  \label{Eq:4pointSimple}
 \end{eqnarray}
For higher order derivatives, it becomes increasingly cumbersome to simplify the expressions explicitly, to check their symmetry property. Nevertheless the full result is explicitly determined by Eq.~\eqref{Eq.final-npoint}.

To verify our general expressions, we have compared against various simple cases where the derivatives can be computed directly with \textsc{Mathematica}. This included the SM Higgs sector and a toy model with an ${\rm SU}(2)$ gauge field, plus  a Weyl fermion doublet, a Weyl fermion singlet and a scalar doublet. We also performed various checks of the one-loop tadpole conditions and mass-squared matrices in the scalar sector of the CP-conserving two Higgs doublet model. 

\section{Examples}\label{sec:examples}
The general result  obtained in the previous sections, Eq.~\eqref{Eq.final-npoint}, can in principle be used for computations involving $N$-point scalar vertices  under the approximation of small external momenta. That is the case, for example, in the construction of effective field theories by the usual matching and running procedure~\cite{Henning:2016lyp} where the heavy degrees of freedom are integrated out.\footnote{
We remind the reader that our calculations assume we are working with a renormalisable theory, thus for the particular case of matching a high scale non-renormalisable effective theory with another effective theory our general formula would have to be generalised.} In principle our result can then be used to obtain all the one-loop matching conditions for a given scalar $N$-point vertex with no derivatives and light particles in the external lines. These are given by computing all diagrams that contain at least one heavy particle in the loop~\cite{Burgess:2007pt,Henning:2016lyp}, which is equivalent to computing all diagrams and subtracting out diagrams involving only light particles, i.e. the matching conditions capture the differences due to the heavy particle interactions. In general, in this procedure, there are contributions from loops with only heavy particles or with a mixture of heavy and light particles~\cite{delAguila:2016zcb,Henning:2016lyp,Ellis:2016enq,Fuentes-Martin:2016uol}. These mixed contributions can be found in the internal sums over the indices corresponding to particles of type~$T$ in Eq.~\eqref{Eq.final-npoint}.

Eq.~\eqref{Eq.final-npoint} can also be used more directly in a phenomenological context to evaluate the one-loop contributions to the effective triple couplings of light scalar states due to heavier degrees of freedom at the electroweak scale. This is interesting because the LHC is now probing the Higgs sector of the SM and the structure of its scalar potential so, in particular, Higgs-to-Higgs decays should play a primary role. We will illustrate this with a real and a complex singlet extension of the SM where one adds a real or a complex scalar field that is a singlet under the SM gauge group. There are many attractive features for this type of models. Namely, they may provide dark matter candidates~\cite{Silveira:1985rk, McDonald:1993ex, Burgess:2000yq,Bento:2000ah,Davoudiasl:2004be,Kusenko:2006rh, vanderBij:2006ne, He:2008qm,Gonderinger:2009jp, Mambrini:2011ik, He:2011gc,Gonderinger:2012rd,Cline:2013gha, Gabrielli:2013hma,Profumo:2014opa}, allow for
electroweak baryogenesis through a strong first-order electroweak phase transition in the early universe~\cite{Menon:2004wv, Huber:2006wf,Profumo:2007wc,Barger:2011vm, Espinosa:2011ax} and they can provide a rich collider phenomenology~\cite{Datta:1997fx, Schabinger:2005ei,BahatTreidel:2006kx, Barger:2006sk, Barger:2007im, Barger:2008jx,O'Connell:2006wi,Gupta:2011gd, Ahriche:2013vqa,Chen:2014ask,Costa:2015llh,Costa:2014qga,Coimbra:2013qq,Robens:2015gla} with Higgs-to-Higgs decays or invisible decays.

The theory and phenomenology of the models that we will consider in the next sub-sections were recently studied in detail in~\cite{Costa:2015llh,Costa:2014qga,Coimbra:2013qq}. We will use, for each model, the samples generated in~\cite{Costa:2015llh}, which contain all the latest phenomenological constraints from: collider experiments (LEP, Tevatron and LHC); electroweak precision observables; dark matter observables (direct detection and relic density upper bounds); and tree level theoretical constraints such as boundedness from below, vacuum stability (the minimum is global) and perturbative unitarity. For further details we refer the reader to~\cite{Costa:2015llh}. We note that in those samples we have used one-loop accurate relations between the input couplings, the Vacuum Expectation Values (VEVs), masses and mixings of the mass eigenstates, with full $p^2$ dependence. These were computed for the CxSM in~\cite{Costa:2015llh} and the details of the calculations for general scalar singlet extension of the SM will appear in~\cite{MOPS2016massGxSMsoon}. In all our calculations we keep all the new scalar sector contributions and only the dominant SM top quark corrections as illustrated below for the triple couplings.

The one-loop corrected couplings, $\lambda_{(S)i_1,\ldots,i_N}$,  for an $N$-point vertex of the light scalar states corrected at one-loop by the heavy states is, under the zero external momentum approximation, given by
\begin{eqnarray}\label{eq:GenEffectCoups}
\lambda_{(S)i_1,\ldots,i_N}&=&\lambda_{(S)i_1,\ldots,i_N}^{{\rm (0)}}+\varepsilon\, \partial_{i_1,\ldots,i_N}^N V^{(1)}_{\rm heavy}\;.
\end{eqnarray}
Here the subscript ``heavy'' indicates that in the indices of the internal sums in the one-loop effective potential at least one index is running over heavy particles, i.e. the contributions with internal sums over only light particles are dropped out. In the remainder of this section, for ease of notation, we will keep the one-loop corrected coupling represented with no superscript, $\lambda_{(S)i_1,\ldots,i_N}$, as in Eq.~\eqref{eq:GenEffectCoups}.  The first term, $\lambda_{(S)i_1,\ldots,i_N}^{{\rm (0)}}$, consists of the tree level diagrams with heavy particles in the internal lines. For the triple couplings that we will obtain in the next section there are no tree level contributions due to the heavy states, only the tree level vertices.

\subsection{The RxSM}
\label{subsec:RxSM}
The simplest model we consider is the real singlet extension of the SM. This is obtained by adding to the SM a real singlet $S$ with a discrete symmetry under $S\rightarrow -S$. The (renormalisable) potential is
\begin{equation}
V_{\rm RxSM}=\dfrac{m^2}{2}H^\dagger H+\dfrac{\lambda}{4}(H^\dagger
H)^2+\dfrac{\lambda_{HS}}{2}H^\dagger H S^2 +\dfrac{m^2_S}{2} S^2 + 
\dfrac{\lambda_S}{4!}S^4 \, ,  \label{eq:V_RxSM}
\end{equation}
 with $m,\lambda,\lambda_{HS},m_S$ and $\lambda_S$ all real. The vacuum of the theory that is consistent with the Higgs mechanism is such that
\begin{equation}\label{eq:vacua:RxSM}
H=\dfrac{1}{\sqrt{2}}\left(\begin{array}{c} G^+ \\
    v+h+ i G^0\end{array}\right) \quad \mbox{and}  \quad S =v_S + s \;,
\end{equation}
where the SM Higgs VEV is $v \approx 246\;\mathrm{GeV}$, and the singlet VEV is $v_S$. This model contains a symmetric phase with $v_S=0$, in which case $S$ is a dark matter candidate, and a broken phase with $v_S\neq 0$, which has two (visible) mass eigenstates $h_1$ and $h_2$ that are  mixtures of $h$ and $S$ and are ordered in mass ($m_{h_1}<m_{h_2}$). We focus on the broken phase, in which the mass eigenstates are given by
\begin{equation}
\left(\begin{array}{c}
h_1\\
h_2
\end{array}\right)
= \left(\begin{array}{cc} \cos\alpha & \sin\alpha\\ -\sin\alpha&\cos\alpha\end{array}\right)
\left(
\begin{array}{c}
h\\
s
\end{array}
\right) \equiv \left(\begin{array}{cc} R_{11} & R_{12}\\ R_{21}&R_{22}\end{array}\right)
\left(
\begin{array}{c}
h\\
s
\end{array}
\right)\, .
\label{Eq:RxSM_mix}
\end{equation}

 In this model, we can evaluate the one-loop contributions to the SM-like Higgs ($h_1\equiv h_{125}$) triple coupling and to the Higgs coupling to a pair of Goldstones (equivalently longitudinal modes of massive vector bosons), in the scenario where it is the lightest of the two scalars, $h_1\equiv h_{125}$, and where the second Higgs, $h_2$, is heavy. We consider scenarios with $h_2>250$~GeV and we also include the top quark contributions in the zero external momentum approximation. This approximation is justified noting that the one-loop contributions due to the top quark appear through a vertex with a Higgs and a pair of top quarks so we expect a kinematic suppression of the $p^2\neq 0$ corrections by $m_{h_1}^2/(4m_t^2)\sim 0.13$. We also consider the heavy scalar mass, $m_{h_2}$, to be larger than $2m_{h_1}\sim 250$~GeV for the same reason, i.e. so that the $p^2\rightarrow 0$ approximation is reliable.

For simplicity let us illustrate the use of Eq.~\eqref{Eq.final-npoint} for the triple $h_{1}$ coupling. At tree level, this is given by~\cite{Costa:2015llh}
\begin{eqnarray}
\lambda_{(S)h_1h_1h_1}^{(0)}&=&\frac{3}{2} \lambda   R_{11}^3 v+3  \lambda_{HS}   R_{11}^2  R_{12}  v_S +3  \lambda_{HS}   R_{11}  R_{12}^2 v+ \lambda_S   R_{12}^3  v_S\; , 
\end{eqnarray}
 The other non-zero tree level scalar couplings that appear in the one-loop calculation are
\begin{eqnarray}
\lambda_{(S)h_1h_1h_2}^{(0)}&=&R_{11}^2 \left(\frac{3 \lambda   R_{21} v}{2}+ \lambda_{HS}   R_{22}  v_S \right)+R_{12}^2 ( \lambda_{HS}   R_{21} v+ \lambda_S   R_{22}  v_S )+\nonumber\\
&& +2  \lambda_{HS}   R_{11}  R_{12} ( R_{21}  v_S + R_{22} v)\;,\nonumber\\
&&\nonumber\\
\lambda_{(S)h_1h_2h_2}^{(0)}&=& R_{11} \left(\frac{3}{2} \lambda   R_{21}^2 v+2  \lambda_{HS}   R_{21}  R_{22}  v_S + \lambda_{HS}   R_{22}^2 v\right)+ \\
&& +R_{12} \left( \lambda_{HS}   R_{21}^2  v_S +2  \lambda_{HS}   R_{21}  R_{22} v+ \lambda_S   R_{22}^2  v_S \right)\;,\nonumber\\
&&\nonumber\\
\lambda_{(S)h_1h_1h_2h_2}^{(0)}&=& R_{11}^2 \left(\frac{3 \lambda   R_{21}^2}{2}+ \lambda_{HS}   R_{22}^2\right)+ R_{12}^2 \left(\lambda_{HS}   R_{21}^2+ \lambda_S   R_{22}^2\right)+4  \lambda_{HS}   R_{11}  R_{12}  R_{21}  R_{22}\; .\nonumber
\end{eqnarray}
To include the dominant top-quark coupling ($y_t$) contributions we also need the following effective fermionic couplings
\begin{eqnarray}
\lambda^{(0)}_{(F)IJh_1}&=&R_{11} v y_t\delta_{IJ}\;,\\
\lambda^{(0)}_{(F)IJh_1h_1}&=&R_{11}^2 y_t^2\delta_{IJ} \; ,
\end{eqnarray}
with $I,J=1,\ldots,6$, i.e. running over three colours and two helicities for the top quark. Finally, applying Eq.~\eqref{Eq:3pointSimple} with at least one heavy particle in the sum we obtain the one-loop correction: 
\begin{eqnarray}
&&\partial_{h_1h_1h_1}^3 V^{(1)}_{\rm heavy}\nonumber\\
 &=&3f_{(S)h_1h_1h_2}^{(1)}\lambda_{(S)h_1h_1h_1}^{(0)}\left(\lambda_{(S)h_1h_1h_2}^{(0)}\right)^2+3f_{(S)h_1h_2h_2}^{(1)}\lambda_{(S)h_1h_2h_2}^{(0)}\left(\lambda_{(S)h_1h_1h_2}^{(0)}\right)^2+\nonumber\\
&&+f_{(S)h_2h_2h_2}^{(1)}\left(\lambda_{(S)h_2h_2h_2}^{(0)}\right)^3+\left(f_{(S)h_2h_2}^{(1)}-k_T+\tfrac{1}{2}\right)\lambda_{(S)h_1h_1h_2h_2}^{(0)}\lambda_{(S)h_1h_1h_2}^{(0)}+\nonumber\\
&&-6 R_{11}^3y_t^4v\left[2v^2y_t^2f_{(F)ttt}^{(1)}+3\left(f_{(F)tt}^{(1)}-k_T+\tfrac{1}{2}\right)\right]\; .
\end{eqnarray}
Note that $f_{(F)ttt}^{(1)}, f_{(S)h_2h_2h_2}^{(1)},\ldots$ are the loop functions defined in Eq.~\eqref{eq:fT-tensors}  for $k=1$, with the indices in subscript ($h_i$ or $t$) denoting which masses are used to evaluate them. Note also that the top quark contribution takes the same form as for the SM Higgs multiplied
by a factor of $R_{11}$ for each $h_1$ in the external legs. This is expected in this type of scalar
singlet extensions of the SM. This comes from the fact that the coupling of each $h_i$ to SM particles is simply
suppressed by a factor given by the overlap of the SM Higgs doublet fluctuation $h$ with the corresponding mass eigenstate -- see Eqs.~\eqref{eq:vacua:RxSM} and~\eqref{Eq:RxSM_mix}.

Similarly, we can obtain a simple expression for the one-loop correction for the $h_1GG$ triple coupling. The result is expressed through the following tree level vertices (here $G$ denotes one of the three real Goldstone degrees of freedom):
\begin{eqnarray}
\lambda_{(S)h_iGG}^{(0)} &=& \frac{1}{2}\lambda   R_{i1} v+ \lambda_{HS}   R_{i2}  v_S  \nonumber\\ 
\lambda_{(S)h_ih_jGG}^{(0)} &=&\frac{1}{2}\lambda   R_{i1}R_{j1}+ \lambda_{HS}   R_{i2}R_{j2}\;. 
\end{eqnarray} 
Then we have
\begin{eqnarray}
&&\partial_{h_1GG}^3 V^{(1)}_{\rm heavy}\nonumber\\
&=&2f_{(S)h_{1}h_{2}G}^{(1)}\lambda_{(S)h_{1}h_{1}h_{2}}^{(0)}\lambda_{(S)h_{2}GG}^{(0)}\lambda_{(S)h_{1}GG}^{(0)}+\nonumber\\
&&+\left(f_{(S)h_{2}h_{2}G}^{(1)}\lambda_{(S)h_{1}h_{2}h_{2}}^{(0)}+f_{(S)h_{2}GG}^{(1)}\lambda_{(S)h_{1}GG}^{(0)}\right)\left(\lambda_{(S)h_{2}GG}^{(0)}\right)^{2}+\nonumber\\
&&+\frac{3}{4}\left[2\lambda_{(S)h_{1}h_{2}GG}^{(0)}\lambda_{(S)h_{1}h_{1}h_{2}}^{(0)}\left(f_{(S)h_{1}h_{2}}^{(1)}-k_{T}+\tfrac{1}{2}\right)+\right.\nonumber\\
&&+\left.\lambda_{(S)h_{2}h_{2}GG}^{(0)}\lambda_{(S)h_{1}h_{2}h_{2}}^{(0)}\left(\logbar m_{h_{2}}^{2}-k_{T}+\tfrac{3}{2}\right)+2\lambda_{(S)h_{1}h_{2}GG}^{(0)}\lambda_{(S)h_{2}GG}^{(0)}\left(\logbar m_{h_{2}}^{2}-k_{T}+\tfrac{1}{2}\right)\right]\nonumber\\
&&-6R_{11}vy_t^4\left(\logbar m_t^2-k_T+\tfrac{3}{2}\right)\;,
\end{eqnarray}
where again, in the last line, the top quark contribution is the SM-like term multiplied by one $R_{11}$ factor due to the $h_1$ in the external leg. 
For other one-loop corrected vertices the steps to follow are similar, but to avoid overshadowing the discussion we omit the detailed expressions in the next section. In all our examples we work in the $\overline{\rm MS}$ scheme ($k_T=3/2$) and choose the renormalisation scale to be the Higgs boson mass, i.e. $\mu=125$~GeV.

\begin{figure}[tb!]
\centering
\begin{tabular}{cc}
\hspace{-9.5mm}\includegraphics[width=0.425\textwidth,clip=true,trim= 20 20 50 20]{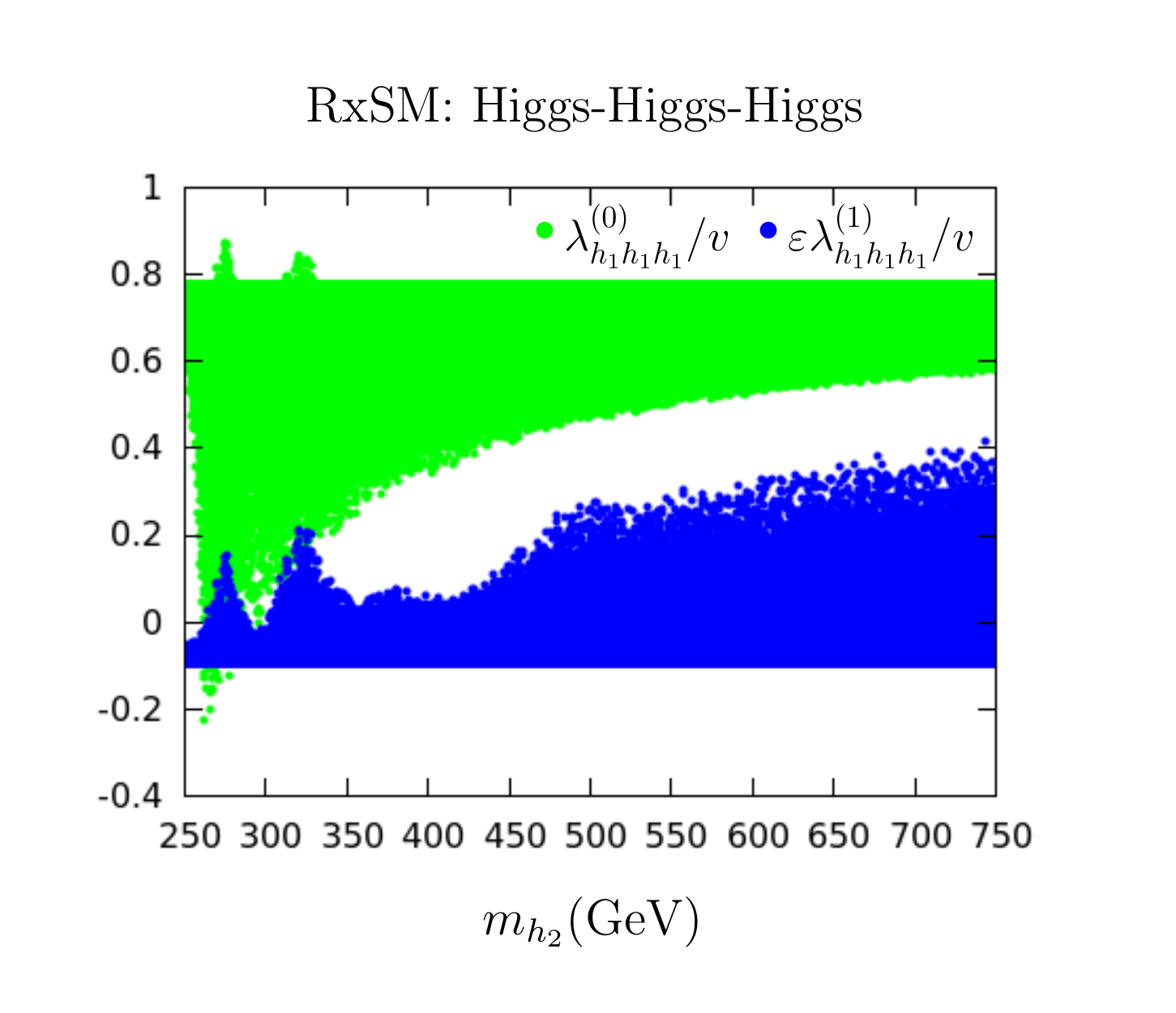}&\hspace{-9.5mm}\includegraphics[width=0.425\textwidth,clip=true,trim=20 20 50 20]{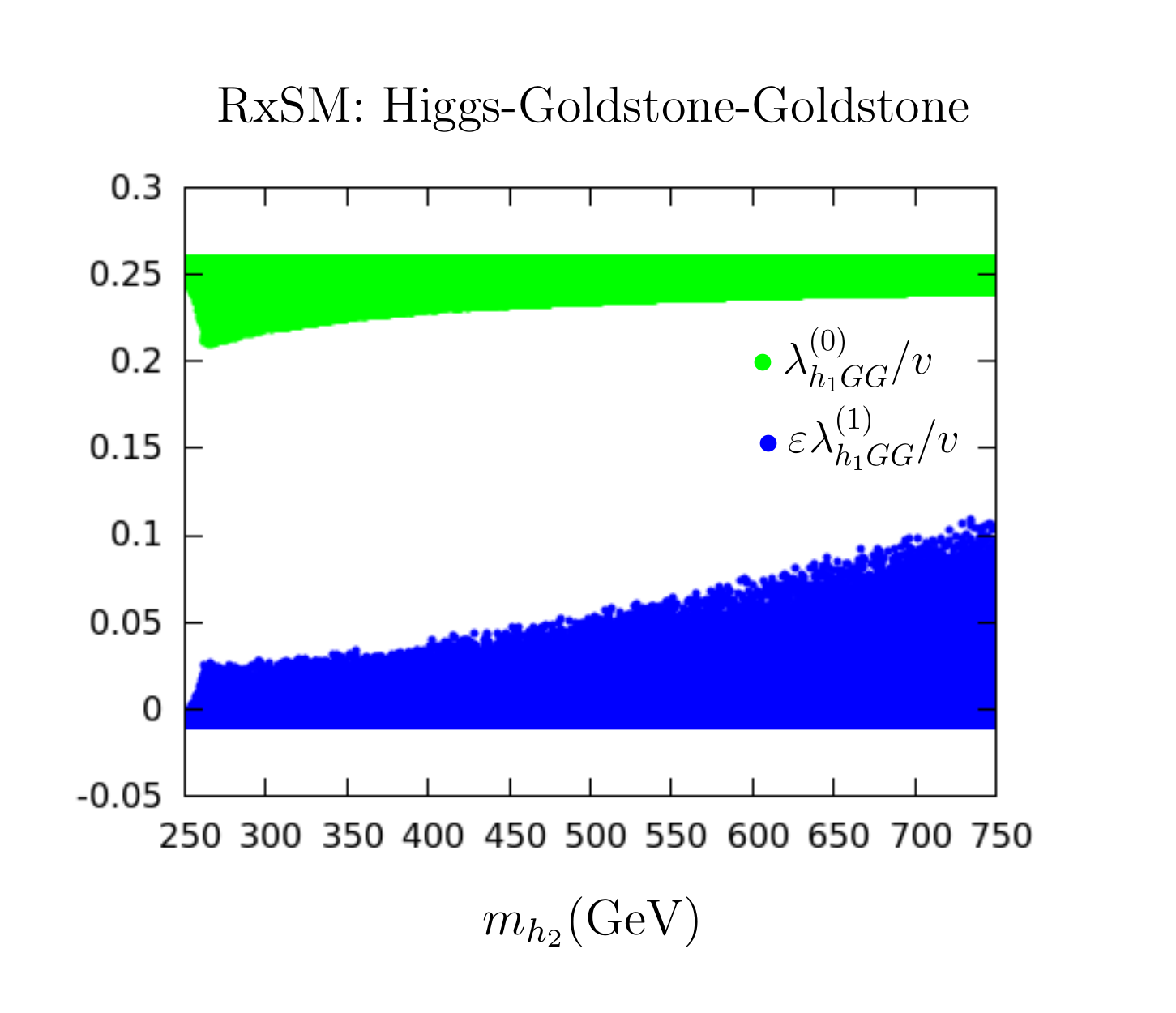}\\
\hspace{-3mm}\includegraphics[width=0.51\textwidth,clip=true,trim= 0 20 0 30]{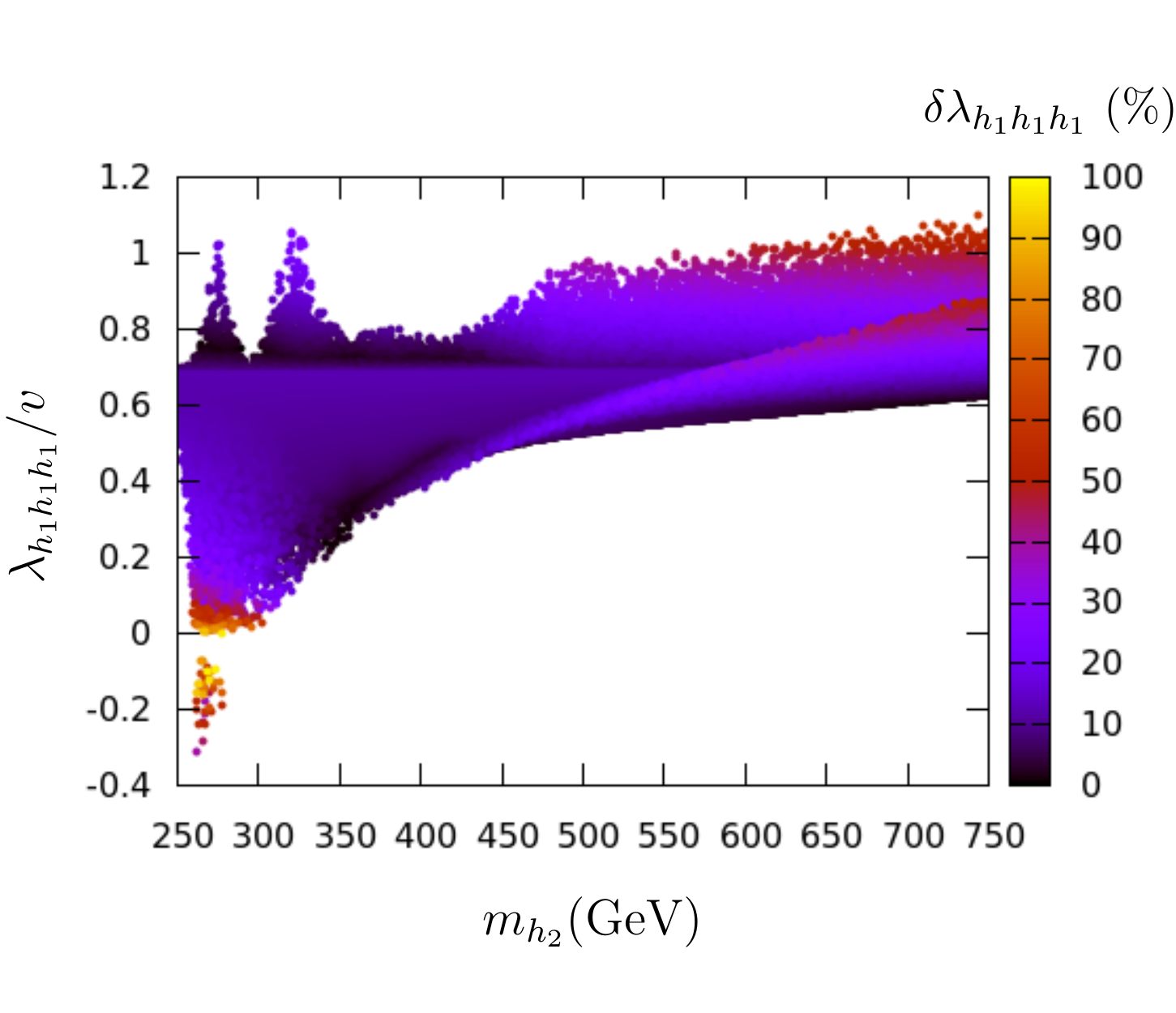}&\hspace{-3mm}\includegraphics[width=0.51\textwidth,clip=true,trim=0 20 0 30]{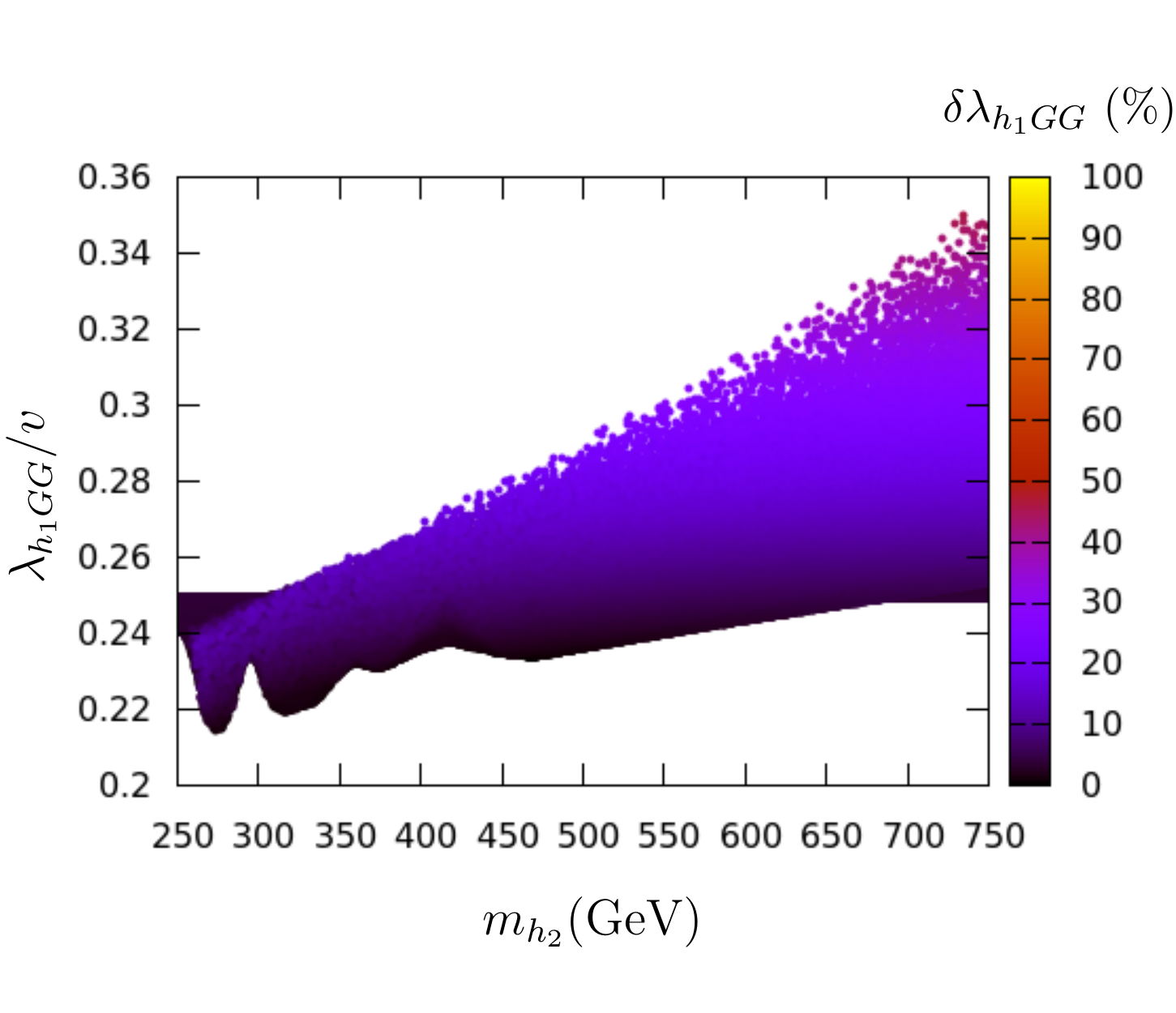}
\end{tabular}
\caption{{\em One loop contributions from a heavy scalar in the RxSM}: The top left panels display the light SM-like Higgs ($h_1$) triple coupling in units of $v$ at tree level (green) and the one-loop correction due to the heavy Higgs ($h_2$) as a function of its mass (blue). In the right hand side panel the colour code is the same except that we display the corresponding quantities for the Higgs-Goldstone-Goldstone coupling. In the bottom panels, we present the same points with the one loop corrected couplings on the vertical axis. The colour scale is the percentage of shift due to the one loop corrections to the couplings.}
\label{fig:RxSM_Hheavy}
\end{figure}

In Fig.~\ref{fig:RxSM_Hheavy} we present a sample of points from the parameter space scan of this model generated in~\cite{Costa:2015llh}, with all the constraints applied as discussed in the beginning of Sect.~\ref{sec:examples}. We show the two triple couplings for the light degrees of freedom in units of $v$ as a function of the mass of the new heavy scalar ($m_{h_2}$). On the left panels we have the SM-like Higgs triple coupling and on the right we have the Higgs-Goldstone-Goldstone coupling, which corresponds to the coupling of the Higgs to a pair of longitudinally polarised vector bosons. The top panels show two layers. In green we display the tree level value of the triple coupling and in blue the one-loop correction. The bottom panels show the one-loop corrected triple couplings with a colour code showing the relative magnitude of the one-loop contribution. 

On the bottom left panel we see that the corrections to the Higgs triple coupling in most of the parameter space are typically\footnote{Note, however, that there are plenty of points with much smaller corrections because the colour scale is such that lighter points are on top of darker points.} $\lesssim 20\%$, except for a region for larger masses $m_{h_2}\gtrsim 500$~GeV where the one-loop corrections can become of the order of\footnote{Note also that we have cut out from the bottom left panel of Fig.~\ref{fig:RxSM_Hheavy} points where $\lambda_{h_1h_1h_1}/v\rightarrow$ is small because the definition of the relative shift $\delta_{h_1h_1h_1}$ becomes singular.} $\sim 50\%$. A future precise measurement of the Higgs triple coupling through Higgs pair production will constrain the deviations of this coupling from the SM value. Thus, in the high mass regions, it will be particularly important to take into account the one-loop corrections due to the new heavy scalar singlet state. Note, however, that the Higgs pair production process is rather small in the SM~\cite{Baglio:2012np} and even at the high luminosity stage of the LHC it will be challenging to measure it very accurately~\cite{CMS:2015nat}. Nevertheless the experimental collaborations are performing dedicated searches and these will provide increasingly tighter bounds for this process -- see e.g.~\cite{ATLASboundHpair}.

The bottom right panel shows a similar pattern for the Higgs coupling to longitudinally polarised vector bosons. For $m_{h_2}\gtrsim 500$~GeV there are scenarios where the corrections to the Higgs-Vector-Vector triple couplings can be larger than $25\%$. Thus we would expect that improved measurements of the off-shell decay of the SM-like Higgs to vector bosons would again constrain the contribution from the heavy scalar singlet and, consequently, the parameter space of this model. 

\subsection{The CxSM}
\label{subsec:CxSM} 
The theory and phenomenology of the complex singlet extension of the SM has recently been studied in detail in~\cite{Costa:2015llh,Costa:2014qga,Coimbra:2013qq}. In this model the Lagrangian of the SM is extended only in the Higgs sector by a complex singlet field $\mathbb{S}=S+i A$. Its scalar potential is
\begin{multline}
V_{\rm CxSM}=\dfrac{m^2}{2}H^\dagger H+\dfrac{\lambda}{4}(H^\dagger H)^2+\dfrac{\delta_2}{2}H^\dagger H |\mathbb{S}|^2+\dfrac{b_2}{2}|\mathbb{S}|^2+
\dfrac{d_2}{4}|\mathbb{S}|^4+\left(\dfrac{b_1}{4}\mathbb{S}^2+a_1\mathbb{S}+{\rm c.c.}\right)
\, ,  \label{eq:V_CxSM} 
\end{multline}
where all parameters are real and the terms in parenthesis softly break a ${\rm U}(1)$ symmetry of the other terms. The doublet and the complex singlet are, respectively, 
\begin{equation}\label{eq:vacua_CxSM}
H=\dfrac{1}{\sqrt{2}}\left(\begin{array}{c} G^+ \\
    v+h+iG^0\end{array}\right) \quad \mbox{and} \quad
\mathbb{S}=\dfrac{1}{\sqrt{2}}\left[v_S+s+i(v_A+ a)\right] \;,
\end{equation}
with $v\approx 246\;\mathrm{GeV}$ the SM Higgs VEV, and $\{v_S, v_A\}$ respectively the VEVs of the real and imaginary parts of the complex singlet. The potential in Eq.\eqref{eq:V_CxSM} is $\mathbb{Z}_2$ symmetric under $A\rightarrow -A$. As a consequence, there are two possible minima that break electroweak symmetry consistently with the Higgs mechanism. If $v_S\neq 0, v_A=0$ then $h,s$ mix into a pair of scalars that are visible at colliders and $A$ does not couple to the other SM-particles, so it is a dark candidate. If both $v_S\neq 0, v_A\neq 0$ then $h,s,a$ all mix and we have three scalars that are visible at colliders (one of them being the observed SM-like Higgs). We focus only on the dark matter phase of this model where we denote the masses of the visible Higgs bosons by $m_{h_1}$ and $m_{h_2}$, and the mass of the dark matter candidate by $m_D$.

 In the case of the dark phase of the singlet extension of the SM we can have a new ingredient in the light spectrum of the theory. We consider the scenario where, again, $m_{h_2}>250$~GeV, $m_{h_1}=125$~GeV is the SM-like Higgs boson, but now there is a dark matter candidate that we choose to be light, $m_D<90$~GeV. Thus, besides the Higgs triple coupling, the one-loop corrections due to the heavy state $h_2$ now affect the Higgs-dark-dark triple coupling. For this vertex, the one-loop corrections are only due to the scalar sector since the dark particle does not couple, at tree level, with the SM particles.  This coupling is very important for dark matter observables, namely the calculation of the relic density that is left over after freeze out in the early Universe, the cross-section for direct detection in underground experiments and invisible decays at colliders. The former two severely constrain the parameter space allowed for the model.
\begin{figure}[bt!]
\centering
 {\small Higgs-Higgs-Higgs [CxSM dark phase, $h_{125}\equiv h_1$]} \vspace{-2.1mm}\\
\mbox{\includegraphics[width=0.506\textwidth,clip=true,trim= 0 20 54 30]{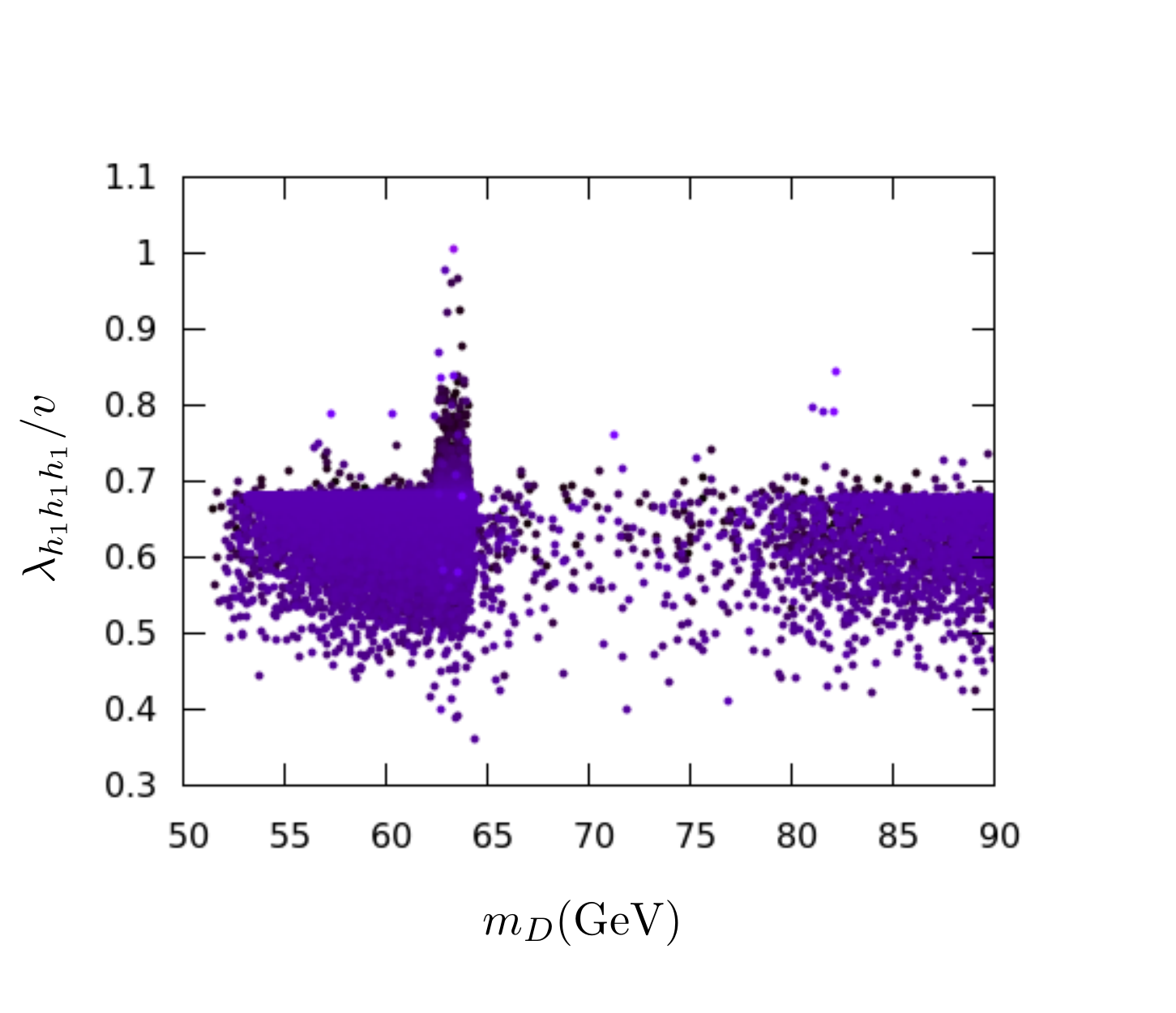}\hspace{1.5mm}\includegraphics[width=0.501\textwidth,clip=true,trim=57.2 20 0 30]{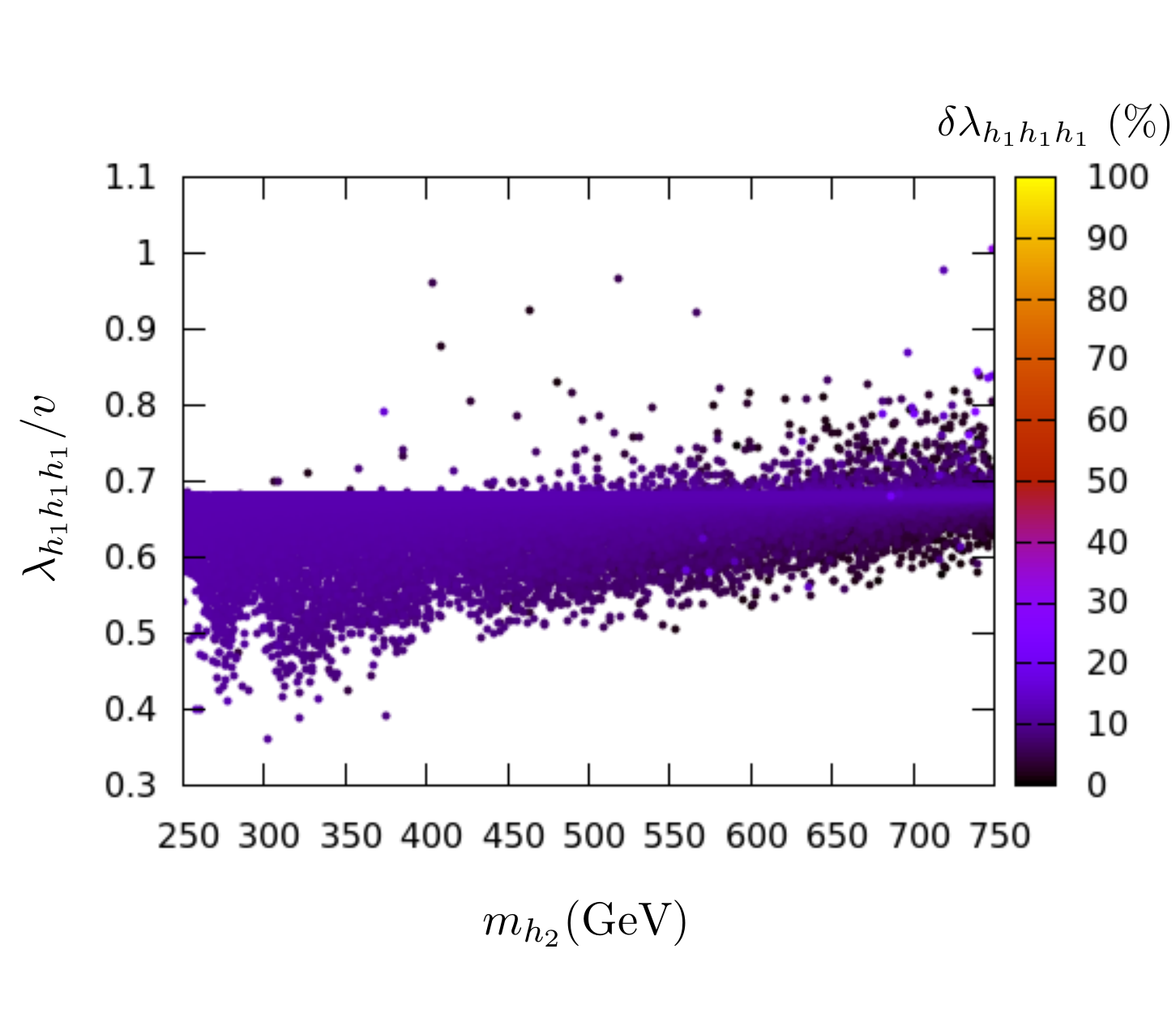}} \vspace{-2mm}\\
 {\small Higgs-Dark-Dark [CxSM dark phase, $h_{125}\equiv h_1$]} \vspace{-2mm}\\
\mbox{\includegraphics[width=0.506\textwidth,clip=true,trim= 0 20 52 30]{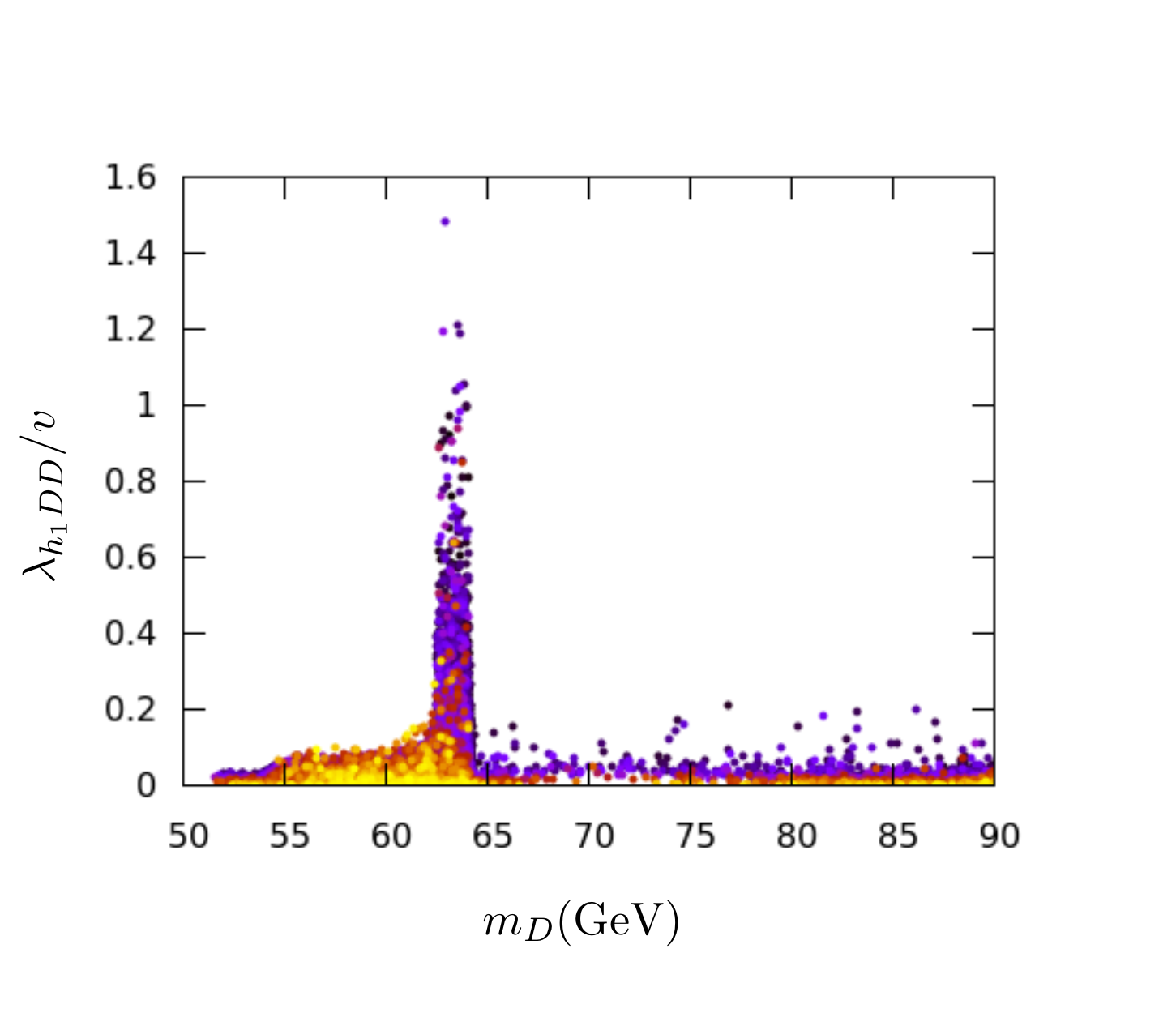}\hspace{1mm}\includegraphics[width=0.496\textwidth,clip=true,trim=57 20 0 30]{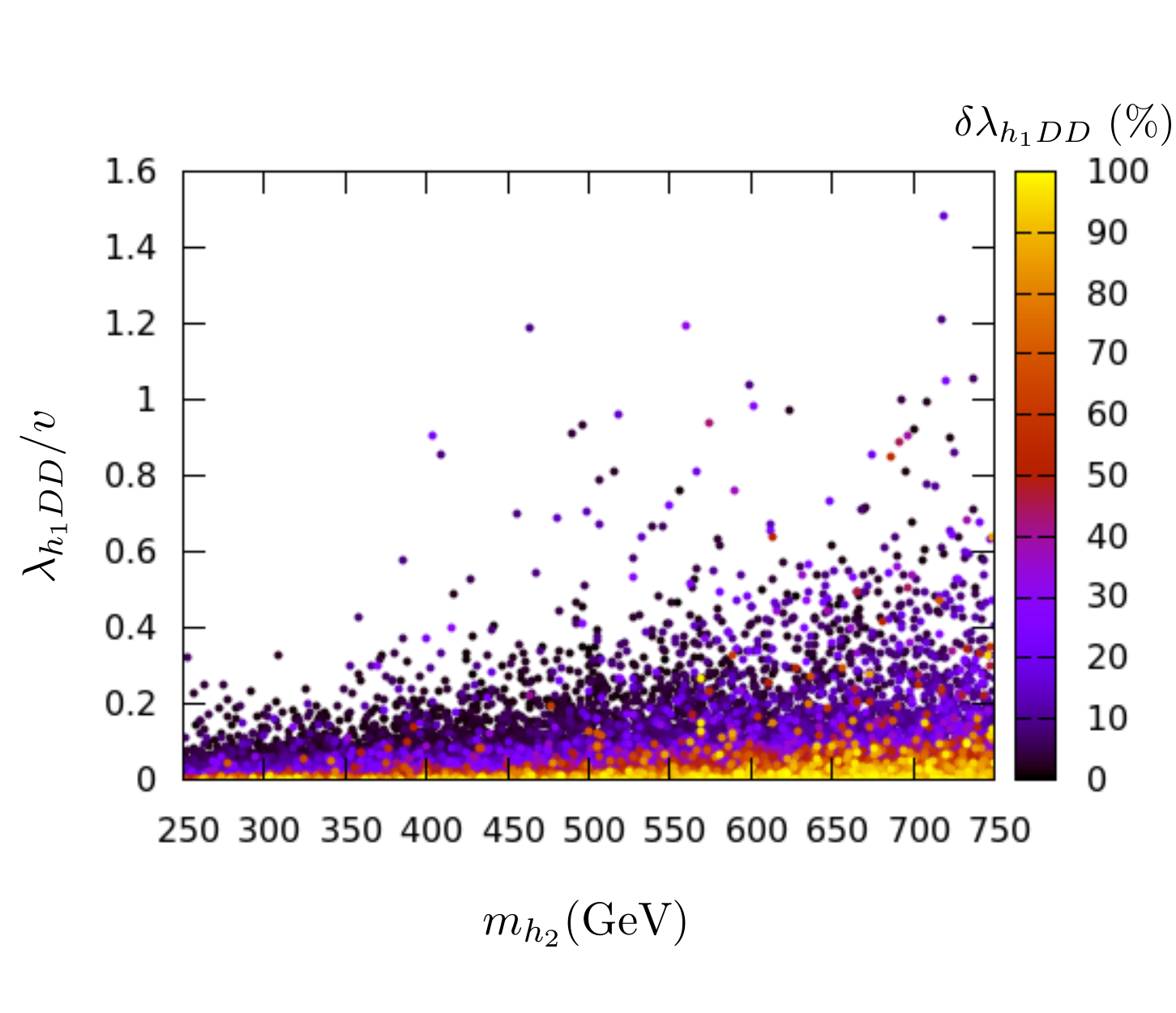}} \\
\caption{{\em One loop contributions from a heavy scalar in the dark phase of the CxSM}: The panels show several projections with triple couplings on the vertical axis versus the dark matter particle mass, $m_D$  (left column) or the heavy scalar mass $m_{h_2}$ (right panel). The colour code on each row is the same for the left and the right panels as well as the triple coupling on the vertical axis.}
\label{fig:CxSM_dark_Hheavy}
\end{figure} 

In Fig.~\ref{fig:CxSM_dark_Hheavy} we present a sample of points from the parameter space scan of this model generated in~\cite{Costa:2015llh}, with all the constraints applied as discussed in the beginning of Sect.~\ref{sec:examples}. The top panels are for the Higgs triple coupling, $\lambda_{h_1h_1h_1}/v$ , and the bottom panels for the Higgs-dark-dark triple coupling $\lambda_{h_1DD}/v$. We first note that in the left panels there is a peak around $m_D\sim m_{h_1}/2$ corresponding to the kinematic region where dark matter annihilates efficiently into Higgs bosons in the early Universe, so the triple couplings are allowed to be larger while the relic density does not become larger than the combination of the observational values from the WMAP and Planck satellites~\cite{Ade:2013zuv,Hinshaw:2012aka}. The loss in density to the right of this threshold is also because of the LUX dark matter direct detection bounds~\cite{Akerib:2013tjd}, which are stronger in the region $\sim 20$~GeV to $~\sim 70$~GeV but become progressively less restrictive outside this window. 

We observe in the top panel that all scenarios in the scan have a one-loop contribution to the Higgs triple coupling that is typically $\lesssim 20\-- 30\%$. In the bottom panel, however, we see that the correction to the Higgs-dark-dark coupling can become as large as the tree level one. This typically happens for points where the tree level contribution is already small, so that the one-loop contribution can enhance it. Since this sample of points was generated with tree level dark matter observables, the potentially large one-loop corrections can exclude such parameter space points or bring back other points excluded by the tree level analysis.

\section{Conclusions}
\label{sec:conclusions}

In this article we have obtained the derivatives of the one-loop CW effective potential with respect to the scalar fields up to an arbitrary number of derivatives. Our central result, Eq.~\eqref{Eq.final-npoint}, was derived for an arbitrary renormalisable quantum field theory in four space-time dimensions. We have first found that it is possible to obtain exact expressions for such derivatives, in a closed form, in the tree level mass eigenbasis. This was done using the CW effective potential in mass independent renormalisation schemes in the Landau gauge. The general result follows from the properties of the matrix-log and suitable combinatorics. It remains valid in an arbitrary basis after applying suitable basis transformations.

Our result can in principle be used to compute effective field theory scalar operators without derivatives with any number of light scalar external legs when integrating out heavy degrees of freedom. It can also be used directly to compute one-loop corrections due to heavy states to scalar interaction vertices. In this context we have analysed two scalar singlet extensions (the RxSM and the dark phase of the CxSM) to observe the effect of a new heavy scalar on the low energy phenomenology of the SM-like 125~GeV Higgs boson and of a light dark matter candidate. We have found that in some cases the corrections to the SM-like Higgs self-coupling and the coupling to vector bosons can be comparable to the tree level value. This shows the importance of computing the full one-loop corrections for these models. Most importantly, the corrections to the coupling between the SM-like Higgs and the dark matter candidate in the dark phase of the CxSM can be large. This can take some scenarios out of the phenomenological viable region, due to one-loop shifts to the dark matter observables that are already constrained, or bring back scenarios that were excluded in a tree level analysis. 

Finally, it would be interesting to study whether our general analytic result could be extended to the derivatives of the two-loop effective potential or if it could be generalised to non-renormalisable theories. 

\section*{Acknowledgements}
A. P. M. and M.S. acknowledge the THEP group at Lund University, where part of this work was developed, for all support and hospitality. J. E. C.-M., R. P. and J. W. acknowledge the warm hospitality of the Gr@v group at Aveiro University.
 The work by J. E. C.-M. was supported by the Crafoord Foundation. A. P. M. and M.S. are funded by FCT through the grants SFRH/BPD/97126/2013 and \mbox{SFRH/BPD/ 69971/2010} respectively.  R. P. and J. W. were partially supported by the Swedish Research Council, contract number 621-2013-428. The work in this paper is also supported by the CIDMA project UID/MAT/04106/2013. 

\appendix

\section{All $N$th order derivatives of the matrix-$\logbar$ term}
\label{app:Matrix_Log}

In Sec.~\ref{sec:OneLoopNpoint} we have summarised the ingredients needed to obtain the zero external momenta contributions to the scalar $N$-point functions from the derivatives of the effective potential. Here we provide details on how to obtain the general result for the $\delta^{(N)}_{a b i_1\ldots i_N}$ tensors from the derivatives of the matrix-$\logbar$. It is indeed instructive to start by deriving the lower order cases first, on one hand because $N=1,2$ are special, and on the other hand because it is then easier to observe the pattern to generalise it to higher orders. For notational simplicity, to avoid carrying around factors of $\mu^2$, we use units of mass such that $\mu^2=1$. In this case $\logbar$ is equivalent to the $\log$ function. The final result will remain valid with the $\mu^2$ dependence simply appearing in masses inside the $\logbar$ factors.  
\subsection{$N=1$}
The quantity we seek is (suppressing for convenience the indices of $\Lambda_{(T)}$, i.e. using matrix notation) 
\begin{eqnarray}
\partial_i\logbar \Lambda_{(T)}&=&-\partial_i\sum_{n=1}^{+\infty}\dfrac{(-1)^n}{n}\left(\Lambda_{(T)}-\mathbf{1}\right)^{n} \nonumber\\
&=&\sum_{n=1}^{+\infty}\dfrac{(-1)^{n+1}}{n}\sum_{q=1}^{n}\left(\Lambda_{(T)}-\mathbf{1}\right)^{q-1}\stackrel{\phantom{K}^{q^{th}term}}{\partial_{i}\Lambda_{(T)}}\left(\Lambda_{(T)}-\mathbf{1}-\right)^{n-1-q}  \nonumber\\
&=&\sum_{n=1}^{+\infty}\dfrac{(-1)^{n+1}}{n}\sum_{q=1}^{n}\left(\Lambda_{(T)}-\mathbf{1}\right)^{q-1}\stackrel{\phantom{K}^{q^{th}term}}{\Lambda_{(T,3)i}}\left(\Lambda_{(T)}-\mathbf{1}\right)^{n-1-q} \; .
\end{eqnarray}
Here we have used the matrix series representation of the matrix log and defined the set of matrices $[\Lambda_{(T,3)i}]_{ab}\equiv \Lambda_{(T)abi}$, which are the cubic vertices of field of type $T$ in the $\Lambda$-basis. Note that on the second line we have obtained $n$ terms by applying the Leibniz rule for the derivative operator acting on the product of $n$ equal terms. Each term is obtained by replacing $\left(\Lambda_{(T)}-\mathbf{1}\right)\rightarrow \partial_{i}\Lambda_{(T)}$ in the the term appearing in position $q$ of the product (named $q^{th}term$). Using the definition of the $\delta^{(N)}_{abi_1\ldots i_N}$ tensors, Eq.~\eqref{eq:N1N2N3N4}, and inserting the orthogonality/unitarity relations for the transformation matrices $\mathcal{U}_{(T)}$ several times between products of matrices\footnote{For a lighter notation, from here on, we use ${\bf m_{(T)}^2}$ to denote the diagonal mass-squared matrix $ {\rm diag}\{m_{(T)}^2\}$.}, we have
\begin{eqnarray}
\delta^{(1)}_{(T)a b i}&=&\sum_{n=1}^{+\infty}\dfrac{(-1)^{n+1}}{n}\sum_{q=1}^{n}\left[\left({\bf m_{(T)}^2}-\mathbf{1}\right)^{q-1}\right]_{a}^{\phantom{a}c}\stackrel{\phantom{K}^{q^{th}term}}{\lambda_{(T)c\phantom{d}i}^{\phantom{(T)c}d}}\left[\left({\bf m_{(T)}^2}-\mathbf{1}\right)^{n-q}\right]_{db} \nonumber\\
&=&\lambda_{(T)abi}\sum_{n=1}^{+\infty}\dfrac{(-1)^{n+1}}{n}\sum_{q=1}^{n}\left(m_{(T)a}^2-1\right)^{q-1}\left(m_{(T)b}^2-1\right)^{n-q} \nonumber\\
&=&\lambda_{(T)abi}f^{(0)}_{(T)ab} \; . \label{eq:N1case}
\end{eqnarray}
In the last line we have used the $N=1$ case of the following general formula
\begin{eqnarray}
&&f\left[x_{1},\ldots,x_{N}\right] \nonumber\\
&\equiv&\sum_{n=1}^{+\infty}\dfrac{(-1)^{n+1}}{n}\sum_{P_{1}=0}^{n-N}\sum_{P_{2}=0}^{n-N-P_{1}}\ldots\sum_{P_{N-1}=0}^{n-N-\sum_{i=1}^{N-1}P_{i}}\left(x_{1}-1\right)^{P_{1}}\left(x_{2}-1\right)^{P_{2}}\ldots\left(x_{N}-1\right)^{n-N-\sum_{i=1}^{N}P_{i}}\nonumber\\
&=&\sum_{i=1}^{N}\dfrac{\logbar x_{i}}{\prod_{k\neq i}\left(x_{i}-x_{k}\right)} \; ,
\end{eqnarray}
which relates to the $f^{(0)}_{(T)abi_1\ldots i_N}$ tensors defined in Eq.~\eqref{eq:fT-tensors}.
\subsection{$N=2$}
In this case, the two derivatives will either produce two cubic vertices or a quartic as follows
\begin{eqnarray}
\partial_{ij}\logbar \Lambda_{(T)}&=&\partial_{ij}\sum_{n=1}^{+\infty}\dfrac{(-1)^{n+1}}{n}\left(\Lambda_{(T)}-\mathbf{1}\right)^{n} \nonumber\\
&=&\sum_{n=1}^{+\infty}\dfrac{(-1)^{n+1}}{n}\left[\sum_{q=1}^{n}\left(\Lambda_{(T)}-\mathbf{1}\right)^{q-1}\stackrel{\phantom{K}^{q^{th}term}}{\partial_{ij}\Lambda_{(T)}}\left(\Lambda_{(T)}-\mathbf{1}\right)^{n-q}+\right. \\ 
&+&\sum_{p=1}^{n-1}\sum_{p<q\leq n}\left(\Lambda_{(T)}-\mathbf{1}\right)^{p-1}\stackrel{\phantom{K}^{p^{th}term}}{\partial_{i}\Lambda_{(T)}}\left(\Lambda_{(T)}-\mathbf{1}\right)^{q-p-1}\stackrel{\phantom{K}^{q^{th}term}}{\partial_{j}\Lambda_{(T)}}\left(\Lambda_{(T)}-\mathbf{1}\right)^{n-q}+  \nonumber \\
&+&\left.\sum_{p=1}^{n-1}\sum_{p<q\leq n}\left(\Lambda_{(T)}-\mathbf{1}\right)^{p-1}\stackrel{\phantom{K}^{p^{th}term}}{\partial_{j}\Lambda_{(T)}}\left(\Lambda_{(T)}-\mathbf{1}\right)^{q-p-1}\stackrel{\phantom{K}^{q^{th}term}}{\partial_{i}\Lambda_{(T)}}\left(\Lambda_{(T)}-\mathbf{1}\right)^{n-q}\right]  \; .\nonumber 
\end{eqnarray}
Clearly, the term with the second derivative is similar to the $N=1$ case up to a replacement of the cubic by a quartic $\lambda_{(T)abij}$  vertex. The last term, however, introduces a new case. Applying again the definition of the $\delta^{(N)}_{abi_1\ldots i_N}$ tensors we have
\begin{eqnarray}
&&\delta^{(2)}_{(T)a b i j}\nonumber\\
&=&\sum_{n=1}^{+\infty}\dfrac{(-1)^{n+1}}{n}\left[\sum_{q=1}^{n}\left[\left({\bf m_{(T)}^2}-\mathbf{1}\right)^{q-1}\right]_{a}^{\phantom{a}c}\lambda_{(T)c\phantom{d}ij}^{\phantom{(T)c}d}\left[\left({\bf m_{(T)}^2}-\mathbf{1}\right)^{n-q}\right]_{db}+\right. \\
&&+\sum_{p=1}^{n-1}\sum_{p<q\leq n}\left[\left({\bf m_{(T)}^2}-\mathbf{1}\right)^{p-1}\right]_{a}^{\phantom{a}c}\lambda_{(T)c\phantom{d}i}^{\phantom{(T)c}d}\left[\left({\bf m_{(T)}^2}-\mathbf{1}\right)^{q-1-p}\right]_{d}^{\phantom{d}e}\lambda_{(T)e\phantom{f}j}^{\phantom{(T)e}f}\left({\bf m_{(T)}^2}-\mathbf{1}\right)^{n-q}_{fb}\nonumber \\
&&+\left.\sum_{p=1}^{n-1}\sum_{p<q\leq n}\left[\left({\bf m_{(T)}^2}-\mathbf{1}\right)^{p-1}\right]_{a}^{\phantom{a}c}\lambda_{(T)c\phantom{d}j}^{\phantom{(T)c}d}\left[\left({\bf m_{(T)}^2}-\mathbf{1}\right)^{q-1-p}\right]_{d}^{\phantom{d}e}\lambda_{(T)e\phantom{f}i}^{\phantom{(T)e}f}\left({\bf m_{(T)}^2}-1\right)^{n-q}_{fb}\right]\;.\nonumber 
\end{eqnarray}
This can be further simplified using the definition of the $f^{(0)}_{(T)abi_1\ldots i_N}$ tensors
\begin{eqnarray}
&&\delta^{(2)}_{(T)a b i j}\nonumber\\
&=&\lambda_{(T)abij}\sum_{n=1}^{+\infty}\dfrac{(-1)^{n+1}}{n}\left[\sum_{q=1}^{n}\left(m_{(T)a}^2-1\right)^{q-1}\left(m_{(T)b}^2-1\right)^{n-q}\right]+ \\
  &&+\lambda_{(T)a\phantom{d}i}^{\phantom{(T)a}d}\lambda_{(T)b{\phantom{d}}j}^{\phantom{(T)b}d}\sum_{n=1}^{+\infty}\dfrac{(-1)^{n+1}}{n}\left[\sum_{p=1}^{n-1}\sum_{p<q\leq n}\left(m_{(T)a}^2-1\right)^{p-1}\times\right. \nonumber\\
&&\phantom{\lambda_{(T)a\phantom{d}i}^{\phantom{(T)a}d}\lambda_{(T)b{\phantom{d}}j}^{\phantom{(T)b}d}\sum_{n=1}^{+\infty}\dfrac{(-1)^{n+1}}{n}}\left.\times\left(m_{(T)d}^2-1\right)^{q-1-p}\left(m_{(T)b}^2-1\right)^{n-q}\right]+(i\leftrightarrow j) \nonumber \\
&=&f^{(0)}_{(T)ab}\lambda_{(T)ab i j}^{\phantom{(T)ci_{1i_{1}}}}+f^{(0)}_{(T)abc}\left(\lambda_{(T)a\phantom{c}i}^{\phantom{(T)a}c}\lambda_{(T)\phantom{j}b j}^{\phantom{(T)}c}+\lambda_{(T)a\phantom{c}j}^{\phantom{(T)a}c}\lambda_{(T)\phantom{j}b i}^{\phantom{(T)}c}\right) \; . \label{eq:N2case}
\end{eqnarray}
\subsection{$N\geq 3$}
\begin{figure}[tb]
\centering\includegraphics[width=0.8\textwidth,clip=true]{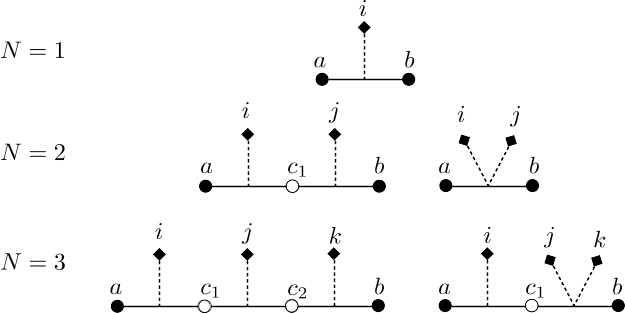}
\caption{{\em Diagrams representing the possible terms appearing in the $\delta^{(N)}_{abi_1\ldots i_N}$ tensors}.}
\label{fig:diagrams_chains}
\end{figure}

From the $N=2$ case, it is now clear that the general result for $N\geq 3$ will be a sum of terms that are generalised versions of the last term we found for $N=2$, Eq.~\eqref{eq:N2case}. This looks like a chain of vertices attached to an $f^{(0)}_{(T)acb}$ tensor (where we have used its total symmetry under exchange of any two indices) with an internal contraction (through the index $c$) and two external indices ${a,b}$ fixed. To obtain the general expression we note the following rules (which follow from applying the differential operator to the series expansion of the matrix-$\logbar$ for general $N$):
\begin{itemize}
 \item For each $N$, the result is a sum over terms containing an $f^{(0)}_{(T)ac_1\ldots c_m b}$ tensor. The allowed values of $m$ are such that\footnote{We denote the integer part of an number $X$ by $[X]$.} $\left[\frac{N-1}{2}\right] \leq m \leq N-1$.
\item Each $f^{(0)}_{(T)a  c_1\ldots c_m b}$ is the base of a chain with two ends fixed ${a,b}$ where we attach (through internal contractions with the $c_i$ indices) cubic or quartic vertices following the pattern 
\begin{equation}\label{eq:PatternChain}
f^{(0)}_{(T)ac_1\ldots c_m b}\times \lambda_{(T)a\phantom{c_1}\{L_1\}}^{\phantom{(T)a}c_1}\lambda_{(T)\phantom{c}\{L_2\}}^{c_2c_3}\ldots \lambda_{(T)\phantom{cccc.}\{L_m\}}^{c_{m-1}c_{m}} \lambda_{(T)\phantom{c_1}b\,\{L_{m+1}\}}^{\phantom{(T)}c_{m}}.
\end{equation}
Each $\{L_k\}$ contains either one index (for cubic vertices) or a pair of indices (for quartic vertices) from the list $\{i_1,\ldots,i_{N}\}$. Note that $m+1$ counts the number of cubic vertices in the chain.
\item The relative position of the internal vertices (without an $a$ or $b$ index) is only relevant for the purpose of counting multiplicities. The internal contractions of such vertices with the $c_i$ indices in $f^{(0)}_{(T)ac_1\ldots c_m b}$ (which is totally symmetric) automatically symmetrises  the result. Thus, it is enough to multiply one representative case by the combinatorial factor under the constraint that the ends of the chain are fixed. Later we can just symmetrise over $\{i_1,\ldots,i_N\}$ to obtain the correct result.
\end{itemize} 
The possible types of terms can be better visualised through graphs as presented in Fig.~\ref{fig:diagrams_chains} for the first three cases. The bottom solid lines with Latin indices $\{a,b,c_i\}$ represent the $f^{(0)}_{(T)ac_1\ldots c_{m}b}$ base chain. The dashed lines represent the $\lambda_{T}$ coupling tensor (one line for cubic or two lines for quartic) with the external scalar indices $\{i,j,\ldots\}$ denoted with diamonds. The external indices $\{a,b\}$ that are uncontracted are denoted with a solid circle, whereas the $c_i$ indices contracted between adjacent $\lambda_{(T)}$ couplings are denoted by open circles.   

The problem then reduces to the computation of combinatorial factors. The natural way of organising the result is to start with the case $m+1=N$ (where all vertices in the chain are cubic) and then at each step (after reducing $m$ by one) we delete a cubic vertex and replace another cubic vertex by a quartic\footnote{This guarantees that the number of $i_k$ indices is always the same.} -- from left to right in Fig.~\ref{fig:diagrams_chains}. This rule is very important since it defines the number of cubic and quartic vertices at each order $n\equiv N-1-m$ in the sum. For the purpose of counting it is in fact convenient to use $n$ instead of $m$. This counts the number of quartic vertices in the chain, so it runs from $n=0$ to $[N/2]$ (simultaneously $m$ runs from $m=N-1$ to $[(N-1)/2]$ ). To obtain the correct multiplicities for the general expression one notes that:
\begin{itemize}
\item At order $n$, the number of ways of distributing the $n$ quartic vertices by the $N-n$ vertices in the chain (before deciding about the positions of the $i_{k}$ indices) is $\binom{N-n}{n}$. As already noted, the relative position of internal vertices in Eq.~\eqref{eq:PatternChain} is only relevant for the purpose of counting multiplicities. Since the internal $c_i$ contractions with the $f^{(0)}_{(T)ac_1\ldots c_m b}$ tensor automatically symmetrise the expression under the internal positions, it is enough to multiply one representative case by the combinatorial factor. The only possibilities (and the respective multiplicites) are as follows:
\begin{enumerate}
\item {\em Two quartic vertices are external -} Here we need to choose  $n-2$ positions out of the remaining $N-n-2$ internal positions (regardless of ordering), to distribute the remaining quartic vertices, i.e. $\binom{N-n-2}{n-2}$.
\item {\em One quartic vertex in the left external position and one cubic in the right one -} In this case, what remains is to choose $n-1$ internal positions out of the remaining $N-n-2$ to distribute the remaining quartic vertices, i.e. $\binom{N-n-2}{n-1}$.
\item {\em Cubic vertex in the left external position and a quartic in the right one -} This is just a flipped case, compared to the previous one, so it has the same multiplicity.
\item {\em Two cubic vertices are external -} In this case we need to distribute all $n$ quartic vertices by $N-n-2$ positions, i.e. $\binom{N-n-2}{n}$.
\end{enumerate}
We note that if we sum the multiplicities of these four cases we obtain $\binom{N-n}{n}$, which is precisely the number of ways of distributing $n$ quartic vertices among $N-n$ available positions. 
\item The final step consists of assigning the $i_{1},\ldots,i_{N}$ indices to the free positions $\{L_k\}$ in each vertex (internal or external). Since the cubic vertices only have one free index to fill, it is easier to use them to start the counting. At order $n$ there are $N-2n$ indices to assign to cubic vertices, so there are $N(N-1)\ldots (N-2n+1)=N!/(2n)!$ ways of distributing them\footnote{Note that now the ordering is relevant. It corresponds to different terms in the expansion of the differential operator (when the Leibniz rule is applied), acting on the polynomials that appear in the series expansion of the matrix-$\logbar$.}. Finally we are left with $2n$ indices to distribute among the $n$ quartic vertices. This corresponds to the number of different ways of arranging the  $2n$ indices in groups of 2. However we must note that in this case the ordering of the two indices within each pair is irrelevant, since the quartic vertices come from a second derivative operator. The combinatorial factor is
\begin{equation}
\binom{2n}{2}\binom{2n-2}{2}\ldots1=\dfrac{(2n)!}{2!(2n-2)!}\times\dfrac{(2n-2)!}{2!(2n-4)!}\times\ldots1=\frac{(2n)!}{2^{n}}\;.
\end{equation}
 Multiplying this factor by the first one, we obtain that the total multiplicity due to the distribution of the indices $\{i_1\ldots i_N\}$ among the various $\{L_k\}$ is \[\frac{N!}{2^{n}}\; .\]
\end{itemize}
Gathering all the multiplicities we can finally write the general expression for $N>2$:
\begin{eqnarray}
\delta^{(N)}_{(T)a b i_1\ldots i_N}&=&{\rm S}_{\left\{ i_{p}\right\} }\sum_{n=0}^{\left[\frac{N}{2}\right]} \frac{N!}{2^{n}}\, f^{(0)}_{(T)ac_{1}\ldots c_{N-1-n}b}\left[\binom{N-n-2}{n}\lambda_{(T)a\phantom{c_{1}}i_{1}}^{\phantom{(T)a}c_{1}}\ldots\lambda_{(T)\phantom{c_{N-}}bi_{N}}^{c_{N-1-n}}\right.+\nonumber\\
&&+\binom{N-n-2}{n-1}\left(\lambda_{(T)a\phantom{c_{1}}i_{1}i_{2}}^{\phantom{(T)}\phantom{a}c_{1}}\ldots\lambda_{(T)\phantom{c_{N-}}bi_{N}}^{c_{N-1-n}}+\lambda_{(T)a\phantom{c_{1}}i_{1}}^{\phantom{(T)a}c_{1}}\ldots\lambda_{(T)\phantom{c_{N-}}bi_{N-1}i_{N}}^{c_{N-1-n}}\right)\nonumber\\
&&+\left.\binom{N-n-2}{n-2}\lambda_{(T)a\phantom{c_{1}}i_{1}i_{2}}^{\phantom{(T)}\phantom{a}c_{1}}\ldots\lambda_{(T)\phantom{c_{ccn}}bi_{N-1}i_{N}}^{c_{N-1-n}}\right]\; . \label{eq:Gen_result}
\end{eqnarray}
Here the internal terms that are suppressed using $\ldots$ are to be replaced by the internal vertices according to the pattern in Eq.~\eqref{eq:PatternChain} in any order\footnote{As explained before the ${\rm S}_{\left\{ i_{p}\right\} }$ takes care of covering all possible cases.}. Also note that whenever the binomial factor is not defined (such that the corresponding chain is inconsistent), the corresponding term is to be deleted. For example for $n=0$, only the first term is defined (it is a term with cubic vertices only).

\bibliographystyle{JHEP}       
\bibliography{references}   

\end{document}